\documentclass[aps,showpacs,showkeys,preprint]{revtex4}
\usepackage{amssymb}
\usepackage{graphicx}
\begin{document}
%\textsf{epsbox}
\preprint{KOBE-TH-04-02}

\title{Protecting the primordial baryon asymmetry in the $SU(2)_{L}$ triplet Higgs model
 compatible with KamLAND and WMAP}



\author{~K.~Hasegawa}
\email[]{kouhei@phys.sci.kobe-u.ac.jp}
\affiliation{Department of Physics, Kobe University, Rokkodaicho 1-1,
Nada ward, Kobe 657-8501, Japan}
%\thanks{}


\date{\today}

\begin{abstract}
We find the condition that the primordial baryon asymmetry is not washed out in the $SU(2)_{L}$
triplet Higgs model by solving the Boltzmann equation. We further require that 
the model is compatible with the recent results of the neutrino oscillation experiments and the 
Wilkinson Microwave Anisotropy Probe, and the constraints on the $\rho$ parameter imposed by 
the CERN LEP. We finally obtain the allowed 
region of the parameters in the model.
\end{abstract}
 


\maketitle

\section{Introduction\label{intro}}
The executed experiments of the neutrino oscillation have made it certain that 
the neutrinos have the very small, but finite masses \cite{S-K, S-K2, Valle-Gon, SNO, SNO2, Chooz}.
The recent KamLAND experiment gave us the more precise information about the neutrino mass matrix 
\cite{kam, kam2,Valle-kam}.
Although the neutrino oscillation experiments are not able to determine the 
magnitudes themselves of the neutrino masses, the recent results of Wilkinson Microwave Anisotropy Probe 
(WMAP) imposed the upper bound of the sum of the neutrino masses \cite{Wmap1,Wmap2,Han,Elg,Wmap3}.
 Since the neutrinos have no masses in the standard model, it is an urgent issue that we extend
 the standard model
and build the model where the neutrinos naturally have the very small masses 
in comparison with the other leptons and quarks. It is expected that the smallness of
the neutrino masses is explained by the Majorana type mass, which only neutrinos
among quarks and leptons are able to have. The Majorana mass terms have the noticeable feature
that they violate the lepton number,  which is conserved at the tree level
in the standard model. A few types of models of the Majorana neutrinos are proposed. 
The seesaw model \cite{Yan,Yan2,Yan3}, the $SU(2)_{L}$ triplet Higgs model \cite{Sch,Laz1}, 
and the Zee model \cite{aZee}
are the three typical models where the standard model is minimally
extended to induce the Majorana masses of the neutrinos within the framework of the $SU(2)_{L} \otimes U(1)_{Y}$
gauge theory.

In this paper we focus on the $SU(2)_{L}$ triplet Higgs model. Especially,
we introduce the model which is built in the framework of the large extra dimension 
and can naturally explain 
the smallness of the neutrino masses by the "shining" mechanism \cite{Ema1, Ema2}.
We live in one brane and the lepton number is violated in the distant other brane.
A messenger field informs us of the lepton number violation in the other brane.
Since the messenger field spreads out in the extra dimensions,
the density of the field is diluted, which is called the "shining" mechanism.
Since we are informed of the only tiny lepton number violation,
the very small neutrino masses are naturally induced in our brane.
Furthermore, based on the 't Hooft's naturalness condition, 
the tiny lepton number violation, that is, the very small neutrino mass,
is stabilized under the quantum corrections.
In this paper we adopt the 4-dimensional effective theory in our brane of the $SU(2)_{L}$ triplet Higgs model
 in the large extra dimension.
 It is also noticed that the $SU(2)_{L}$ triplet Higgs model must not induce 
the dangerous pseudo Majoron which is inconsistent with the accelerator experiments as 
the CERN LEP, and the model is constrained from the observation of the $\rho$ parameter in 
the LEP \cite{gri, erl}. The lepton flavor violation in this model is analyzed in \cite{Chu2, Kak}.

The baryon number in the present universe is well known to be \cite{Wmap1}
\begin{eqnarray}
B_{0}\equiv \frac{n_{B}}{s}=(8.4 \sim 9.1) \  \times 10^{-11}, 
 \end{eqnarray}
where $n_{B}$ is the baryon number density and $s$ is the entropy density.
Some scenarios where the baryon number is dynamically generated from 
vanishing initial baryon number, $B_{ini}=0$, are proposed. 
The GUT scenario \cite{Yos, Barg}, the electroweak scenario \cite{Kuz, Coh}, 
the Fukugita-Yanagida scenario \cite{Fuk}, and the Affleck-Dine scenario \cite{Aff}
are proposed as the typical scenarios of the baryogenesis.
The GUT and electroweak scenarios have already been almost ruled out. 
It is often mentioned that these scenarios of the baryogenesis 
have a double-edged behaver, that is, the scenarios of the baryogenesis simultaneously have a potential to 
erase the existing baryon number. Especially, while the mass models of the Majorana neutrinos 
have a potential to generate the baryon number through the leptogenesis as in the Fukugita-Yanagida scenario, 
these models simultaneously have a potential to erase arbitrary initial lepton and
therefore baryon numbers. There exist the sphaleron-anomaly processes which create and annihilate the baryon 
in an equal amount as that of the lepton in the standard model \cite{Hoo, Man, Kli, Arn, Bod}. 
When the the sphaleron-anomaly processes and the lepton number violating processes in the 
mass models of the Majorana neutrinos are simultaneously in equilibrium, the arbitrary
 initial baryon and lepton numbers are actually washed out. Since the sphaleron-anomaly processes are 
 in equilibrium at the large region of temperature between 100 GeV and
 $10^{12}$ GeV, whether the lepton number violating 
 processes are in equilibrium or not in the region handles the fate of the cosmological baryon number.

In this paper, we assume that the primordial baryon number is generated by
some unknown mechanism at the very early universe and
require that the primordial baryon number is not washed out 
in the mass models of the Majorana neutrinos where the baryon number is 
never generated. The situation where the baryon number is never generated via leptogenesis
in the Majorana mass models is typically realized when we assume that the lepton sector has no 
CP violation because of the Sakharov's three conditions \cite{Sak}. Although we assume no CP violation in the 
following analyses, the results of this paper are valid as long as the mass models of the Majorana neutrinos
are not able to generate the present baryon number. 
Actually, it is shown that even if the $SU(2)_{L}$ triplet Higgs model, which we use in this paper, 
has CP violation in the lepton sector, the baryon number can not be 
generated \cite{Ern}. The leptogenesis in the other extended triplet 
Higgs models is analyzed in \cite{Laz2,Laz3}.
In  order to generate the baryon number via leptogenesis,
two triplet Higgs fields must be introduced.
The similar analyses are carried out in the GUT scenario \cite{Fry1, Fry2}
 and in the Fukugita-Yanagida scenario
\cite{Fuk2, Hav, Nel, Buc3,  Fis, Oli1, Oli2, Dre}.

In order to escape from washing out of the baryon number in the mass models of the Majorana neutrinos,
it is necessary that the lepton number violating processes are out of equilibrium.
An out-of-equilibrium process means that the reaction rate of a process is smaller than the Hubble parameter
at that time. In order to violate the lepton number explicitly in the $SU(2)_{L}$ triplet Higgs model, two kinds 
of interactions must be newly introduced. At first sight, it is not clear whether both of the two kinds of 
interactions must be out of equilibrium or either of the two kinds of interactions must be out of equilibrium
in order to protect the primordial baryon number. The answer we obtain in this paper
is that if either of the two kinds of interactions is out of equilibrium, the primordial baryon number 
is not washed out. If either of the two kinds of interactions is out of equilibrium, the lepton number
is considered not to be explicitly broken and an approximately preserved global $U(1)$ charge which contains
the lepton number exists.
Thus the primordial baryon number is protected in proportion to the initial value of the approximate $U(1)$
charge. It is the first purpose in this paper that we can concretely estimate the condition to protect the 
primordial baryon number by solving the Boltzmann equation in the $SU(2)_{L}$ triplet Higgs model.
We next analyze whether the $SU(2)_{L}$ triplet Higgs model can simultaneously satisfy
the obtained condition to protect baryon number and the results of the neutrino oscillation experiments or not.
We further require that the model satisfies the results of WMAP and the constraints on the $\rho$ parameter.
It is the second purpose in this paper that we require that the model satisfies
all the  above conditions and obtain the allowed region of the parameters in the model.
The similar analyses are carried out in the seesaw model \cite{Has2} 
and in the Zee model \cite{Haba, Has1}.

The outline of this paper is as follows: in Sec. \ref{trimodel}, we review the $SU(2)_{L}$ triplet Higgs model
where the smallness of the neutrino masses is naturally explained. We make sure that 
the dangerous pseudo Majoron which is inconsistent with the accelerator experiments as 
LEP does not appear in the model. In Sec. \ref{Washing}, we obtain the condition to protect the primordial
baryon number by solving the Boltzmann equation in the $SU(2)_{L}$ triplet Higgs model.
In Sec. \ref{triprotect}, we require that the $SU(2)_{L}$ triplet Higgs model satisfies 
the condition to protect the baryon number, the results of the neutrino oscillation experiments and WMAP,
and the constraints on the $\rho$ parameter. We finally obtain the allowed region of the parameters
in the model. Sec. \ref{trisum} is devoted to a summary.

\section{$SU(2)_{L}$ Triplet Higgs Model   \label{trimodel}}
In this section we first introduce the $SU(2)_{L}$ triplet Higgs model in 
the normal four dimensional space-time as an effective theory of
the model in the large extra dimension \cite{Ema1, Ema2}.
It should be noted that two kinds of interactions must be newly introduced
in order to violate the lepton number explicitly.
We second show that the very small neutrino masses are naturally induced and 
simultaneously confirm that the dangerous pseudo Majoron which is inconsistent 
with the accelerator experiments as LEP does not appear 
under the suitably assumed Higgs potential.
We finally check how the observations of the $\rho$ parameter at LEP
constrain the model.

We newly introduce the $SU(2)_{L}$ triplet Higgs fields $\Delta$ assigned with 
the hypercharge $Y=2$  in addition to the fields of the standard model as
 \begin{eqnarray}
\Delta \equiv \left(\begin{array}{cc}
                                    \xi^{+}/\sqrt{2}  &   \xi^{++}    \\
                                     \xi^{0}  & -\xi^{+}/\sqrt{2}  \label{ddef}
									 \end{array}\right).
\end{eqnarray}
The Yukawa interactions of the triplet Higgs with the leptons are written as 
\begin{eqnarray}
 {\cal L}^{yukawa}_{\nu}
 &=& -\frac{1}{2}f^{\alpha \beta}Tr[T_{l^{\alpha},l^{\beta}}\Delta]
 +\mbox{h.c.}  \label{lld} \\ 
 &=& -\frac{1}{2}f^{\alpha \beta}\biggl[ \overline{(\nu^{\alpha})^{c}}\nu^{\beta}\xi^{0}-\frac{1}{\sqrt{2}}
   (\overline{(\nu^{\alpha})^{c}}e^{\beta}
   +\overline{(e^{\alpha})^{c}}\nu^{\beta})\xi^{+}-\overline{(e^{\alpha})^{c}}e^{\beta}\xi^{++}\biggr]
+\mbox{h.c.},  \nonumber
\end{eqnarray}
where $T_{l^{\alpha},l^{\beta}}$ is a $SU(2)_{L}$ triplet which is composed
of the two lepton doublets,
 \begin{eqnarray}
T_{l^{\alpha},l^{\beta}} \equiv \left(\begin{array}{cc}
                                    -\overline{(\nu_{L}^{\alpha})^{c}}e_{L}^{\beta}  & 
									\overline{(\nu_{L}^{\alpha})^{c}}\nu_{L}^{\beta}     \\
                                     -\overline{(e_{L}^{\alpha})^{c}}e_{L}^{\beta}  & 
									 \overline{(\nu_{L}^{\alpha})^{c}} e_{L}^{\beta}
									 \end{array}\right),
\end{eqnarray}
and $f^{\alpha \beta}$ is the symmetric Yukawa coupling constants ($f^{\alpha \beta}=f^{\beta \alpha}$). Since we take
the base where the mass matrix of the charged leptons is diagonalized,
the indices, $\alpha$ and $\beta$ take $e, \mu,$ or $\tau$.
When the neutral triplet Higgs field $\xi^{0}$ gets the vacuum expectation value (VEV),
the interactions in (\ref{lld}), whose mass dimension is four, induce
the Majorana mass terms of the neutrinos, which explicitly violate the lepton number.
However when we assign the triplet Higgs fields 
with the lepton number $L_{\Delta}=+2$, none of the interactions in the model violate 
the lepton number at the symmetric phase
where the neutral triplet Higgs field has no VEV, that is, the lepton number is not explicitly violated.
In the broken phase, neutrinos acquire Majorana masses via 
the spontaneous breakdown of the lepton number, which in turn means that a Majoron, which is inconsistent 
with the accelerator experiments as LEP, appears.
Thus, in order to avoid the dangerous Majoron, we need to violate the lepton number explicitly.
For such purpose, we further introduce the cubic interaction of the triplet Higgs with ordinary Higgs doublet 
$\Phi=(\phi^{+}, \phi^{0})^{t}$,
\begin{eqnarray}
 {\cal L}^{cubic}&=&-\frac{1}{2}A Tr[T_{\Phi,\Phi}\Delta^{\dagger}] +\mbox{h.c.} \nonumber \\
   &=& -\frac{1}{2}A \biggl[ (\phi^{+})^{2}\xi^{--}-\sqrt{2}
 \phi^{+}\phi^{0}\xi^{-}-(\phi^{0})^{2}\xi^{0 \ast}\biggr]+\mbox{h.c.}, \label{ppd5} 
\end{eqnarray}
where $T_{\Phi,\Phi}$ is a $SU(2)_{L}$ triplet which is composed
of the product of the Higgs doublet :
\begin{eqnarray}
T_{\Phi,\Phi} \equiv \left(\begin{array}{cc}
                                    -\phi^{+}\phi^{0}  & \phi^{+}\phi^{+}   \\
                                     -\phi^{0}\phi^{0}  & \phi^{+}\phi^{0}
									 \end{array}\right), \label{phiphi}
\end{eqnarray}
which has the hypercharge $Y=+2$ and the lepton number $L=0$, and $A$ is the coupling constant.

The coupling constant $A$ in (\ref{ppd5}) can be naturally made very small in 
the large extra dimension scenario by the shining mechanism \cite{Ema1, Ema2}. 
Since the neutrino masses are in proportion to the VEV of the neutral triplet Higgs field
and the VEV turns out to be in proportion to the coupling constant $A$, 
the smallness of the neutrino masses is naturally explained
 by the smallness of the coupling constant $A$ in the model.
Here, under a simplified Higgs potential, we show that 
the VEV of the neutral triplet Higgs field is in fact in proportion to the coupling constant $A$
and simultaneously confirm that the 
dangerous pseudo Majoron does not appear. 
We work in a simplified Higgs potential, 
\begin{eqnarray}
V(\Phi, \Delta)=-\mu^{2}\Phi^{\dagger}\Phi+\lambda(\Phi^{\dagger}\Phi)^{2} 
                            +M^{2}Tr[\Delta^{\dagger}\Delta]
							+\frac{1}{2}\bigl(A \ Tr[T_{\Phi,\Phi}\Delta^{\dagger}] +\mbox{h.c.}\bigr),    \label{pot} 
\end{eqnarray}
where $\mu^{2}$ and $M^{2}$ are positive \footnote{In order to be renormalizable, we must 
include all operators which are allowed by the symmetries associated with a model and which
 have the dimension less than five in a bare Lagrangian density. In this case, although we should add
%\begin{eqnarray}
$
\delta V =\lambda_{1} (Tr[\Delta^{\dagger} \Delta])^{2}+\lambda_{2}
Tr[\Delta^{\dagger} \Delta^{\dagger}][\Delta \Delta]+\lambda_{3}
\Phi^{\dagger}\Phi  \ Tr[\Delta^{\dagger}\Delta]+\lambda_{4} 
\Phi^{\dagger} \Delta^{\dagger} \Delta \Phi,
$
%\end{eqnarray}
 to the potential, we assume that the coupling constants $\lambda_{i} (i=1 \sim 4)$ are
 enough small to neglect $\delta V$.
It is shown in \cite{Ema2} that even if we add $\delta V$, the same mass matrix of the neutrinos
is obtained.}.
We extract the sector with electromagnetic charge $Q=0$ in (\ref{pot}),
\begin{eqnarray}
  V(\phi_{1}, \phi_{2}, \xi_{1}, \xi_{2})_{Q=0}=-\frac{\mu^{2}}{2}(\phi_{1}^{2}+\phi_{2}^{2})
+\frac{\lambda}{4}(\phi_{1}^{2}+\phi_{2}^{2})^{2}  
+\frac{M^{2}}{2}(\xi_{1}^{2}+\xi_{2}^{2}) \nonumber \\
-\frac{A}{2\sqrt{2}} \biggl[(\phi_{1}^{2}-\phi_{2}^{2})\xi_{1}+2\phi_{1}\phi_{2}\xi_{2}\biggr],
    \label{pot1}
\end{eqnarray}
where the real fields $(\phi_{1},\phi_{2},\xi_{1},\xi_{2})$ are defined as
\begin{eqnarray}
\phi^{0}\equiv \frac{\phi_{1}+i\phi_{2}}{\sqrt{2}} \ \mbox{and} \ \xi^{0}\equiv  \frac{\xi_{1}+i\xi_{2}}{\sqrt{2}}.
\end{eqnarray}
Solving the conditions for the extremum of the potential (\ref{pot1}),
  \begin{eqnarray}
\frac{\partial V_{Q=0}}{\partial \phi_{1}}=\frac{\partial V_{Q=0}}{\partial \phi_{2}}=
\frac{\partial V_{Q=0}}{\partial \xi_{1}}=\frac{\partial V_{Q=0}}{\partial \xi_{2}}=0. \label{min}
\end{eqnarray}
we  obtain the vacuum expectation values of $\phi_{1}$, $\phi_{2}$,  $\xi_{1}$, and $\xi_{2}$,
  \begin{eqnarray}
  <\phi_{1}>=\frac{\mu}{\sqrt{\lambda-A^{2}/4M^{2}}}, \  <\phi_{2}>=0, 
  \ <\xi_{1}>=\frac{Av^{2}}{2\sqrt{2}M^{2}}, \ \mbox{and} \ <\xi_{2}>=0, \label{vev1}
\end{eqnarray}
and define as $v \equiv <\phi_{1}>$ and $v_{\Delta} \equiv <\xi_{1}>$.
The shifts, $\sigma, N, \sigma_{\Delta}, N_{\Delta}$ from the VEVs (\ref{vev1})
are defined as
 \begin{eqnarray}
 (\phi_{1}, \phi_{2})=(v+\sigma, N) \ \ \mbox{and} \ \ 
 (\xi_{1}, \xi_{2})=(v_{\Delta}+\sigma_{\Delta}, N_{\Delta}).  \label{defv}
\end{eqnarray}
Accordingly, the potential (\ref{pot1}) is rewritten as
\begin{eqnarray}
V(\sigma, N, \sigma_{\Delta}, N_{\Delta} )_{Q=0}&=&-\frac{\lambda}{4}v^{4}-
\frac{A^{2}v^{4}}{16M^{2}}
+\frac{1}{2}(2\lambda v^{2})\sigma^{2}+\frac{1}{2}M^{2}\sigma_{\Delta}^{2}
-\frac{Av}{\sqrt{2}}\sigma\sigma_{\Delta}  \nonumber \\ 
& &+\frac{A^{2}v^{2}}{4M^{2}}N^{2}+\frac{1}{2}M^{2}N_{\Delta}^{2}
-\frac{Av}{\sqrt{2}}NN_{\Delta}, 
\end{eqnarray}
where we have kept only the terms up to quadratic in the fields.
Here we define the potential $V(\sigma, \sigma_{\Delta})$ and $V(N, N_{\Delta})$
as 
\begin{eqnarray}
V(\sigma, \sigma_{\Delta}) &\equiv& \frac{1}{2}(2\lambda v^{2})\sigma^{2}+
\frac{1}{2}M^{2}\sigma_{\Delta}^{2}-\frac{Av}{\sqrt{2}}\sigma\sigma_{\Delta},     \nonumber \\
V(N, N_{\Delta}) &\equiv& \frac{A^{2}v^{2}}{4M^{2}}N^{2}+\frac{1}{2}M^{2}N_{\Delta}^{2}
-\frac{Av}{\sqrt{2}}NN_{\Delta}.  \nonumber
\end{eqnarray}
The potential $ V(\sigma, \sigma_{\Delta})$ is diagonalized by an 
orthogonal transformation 
\begin{eqnarray}
\left(\begin{array}{c}
                                       \sigma^{'} \\
                                        \sigma_{\Delta}^{'}      
                                   	  \end{array}\right)
									   =R(\theta_{\sigma}) 
									   \left(\begin{array}{c}
                                  \sigma \\
                                    \sigma_{\Delta}   
                                   	\end{array}\right),
 \end{eqnarray}
where the rotation matrix $R(\theta_{\sigma})$ is defined as 
\begin{eqnarray}
 R(\theta_{\sigma})\equiv  \left(\begin{array}{cc}
 \cos \theta_{\sigma} & -\sin  \theta_{\sigma} \\
 \sin  \theta_{\sigma} & \cos  \theta_{\sigma}
 \end{array}\right).
 \end{eqnarray}
 Here the rotational angle $\theta_{\sigma}$ and the two mass eigenvalues, 
 $m_{\sigma^{'}}^{2}$ and $m_{\sigma_{\Delta}^{'}}^{2}$ are given as
\begin{eqnarray}
\tan 2\theta_{\sigma} &=&\frac{-4vv_{\Delta}}{v^{2}-2\lambda v^{4}/M^{2}}, \\
m_{\sigma^{'}}^{2}, m_{\sigma_{\Delta}^{'}}^{2} &=&
 \frac{M^{2}+2\lambda v^{2}\pm \sqrt{(M^{2}-2\lambda v^{2})^2+16 M^{2} v_{\Delta}^{2}/v^{2}}}{2},
\end{eqnarray}
where $m_{\sigma^{'}}^{2}$ takes the positive sign.
In a similar way, in order to diagonalize the potential $V(N, N_{\Delta})$,
we rotate the base $(N, N_{\Delta})$ as
\begin{eqnarray}
\left(\begin{array}{c}
                                       N^{'} \\
                                        N_{\Delta}^{'}      
                                   	  \end{array}\right)
									   =  R(\theta_{N}) 
									   \left(\begin{array}{c}
                                  N \\
                                    N_{\Delta}   
                                   	\end{array}\right),
\end{eqnarray} 
where the rotational angle $\theta_{N}$ and the two mass eigenvalues $m_{N^{'}}^{2}$ and 
$m_{N_{\Delta}^{'}}^{2}$ are given as
\begin{eqnarray}
\tan 2\theta_{N} &=& \frac{-4vv_{\Delta}}{v^{2}-4v_{\Delta}^{2}}, \\
m_{N^{'}}^{2}=0  &\mbox{and} & \ \  m_{N_{\Delta}^{'}}^{2}=M^{2}
\Biggl(1+4\frac{v_{\Delta}^{2}}{v^{2}}\Biggr). 
\label{ngs1}
\end{eqnarray}
We next extract the terms with non-zero electromagnetic charge, $Q\not=0$ in (\ref{pot}) : 
\begin{eqnarray}
V_{Q\not=0}=-\mu^{2}|\phi^{+}|^{2}+\lambda(|\phi^{+}|^{4}+2|\phi^{+}|^{2}|\phi^{0}|^{2})
+M^{2}(|\xi^{+}|^{2}+|\xi^{++}|^{2})   \nonumber \\
	 +\frac{1}{2}\Bigl[A\bigl((\phi^{+})^{2}\xi^{--}-\sqrt{2}\phi^{0}\phi^{+}\xi^{-}\bigr)+
	 \mbox{h.c.} \Bigr].
	 \label{cpot}
\end{eqnarray}
Since the potential (\ref{cpot}) induces the mixing term between $\phi^{+}$ and $\xi^{+}$
at the symmetry broken phase, we diagonalize the mixing term by using a rotation,
\begin{eqnarray}
\left(\begin{array}{c}
                                       \phi^{'+}  \\
                                            \xi^{'+}  
                                   	  \end{array}\right)
									   =R(\theta_{+})
									   \left(\begin{array}{c}
                                   \phi^{+} \\
                                      \xi^{+} 
                                   	\end{array}\right).
 \end{eqnarray}
 Here the rotational angle $\theta_{+}$ and the two mass eigenvalues 
 $m_{\phi^{'+}}^{2}$ and $m_{\xi^{'+}}^{2}$ are defined as
\begin{eqnarray}
\tan 2\theta_{+} &=& \frac{-2\sqrt{2}vv_{\Delta}}{v^{2}-2v_{\Delta}^{2}}, \\
m_{\phi^{'+}}^{2}=0  &\mbox{and}& \ \  m_{\xi^{'+}}^{2}=m_{N_{\Delta}^{'}}^{2}=M^{2}
\Biggl(1+4\frac{v_{\Delta}^{2}}{v^{2}}\Biggr).
\label{ngs2}
\end{eqnarray}
From (\ref{ngs1}) and (\ref{ngs2}) we thus confirm that there are no dangerous pseudo Majoron which 
is inconsistent with the experiments.
In fact, $N^{'}$ and $\phi^{'+}$ are would-be N-G bosons, and the masses of remaining physical
Higgs scalar $N_{\Delta}^{'}$ and $\xi^{'+}$ can be kept large 
in the case that $M^{2}$ is large enough, even if $A$ is small.
When the gauge symmetry is spontaneously broken and the neutral Higgs field, $\xi^{o}$ 
gets the VEV $v_{\Delta}/\sqrt{2}$, the Majorana mass terms of the left-handed neutrinos 
are induced in (\ref{lld}) as 
  \begin{eqnarray}
 {\cal L}^{Majorana}=-\frac{1}{2}m^{\alpha \beta}\overline{(\nu_{L}^{\alpha})^{c}}
 \nu_{L}^{\beta}+\mbox{h.c.},  
\end{eqnarray}
where the neutrino mass matrix $m^{\alpha \beta}$ is defined as
\begin{eqnarray}
m^{\alpha \beta} \equiv f^{\alpha \beta}<\xi^{0}>=f^{\alpha \beta}\frac{Av^{2}}{4M^{2}}. 
  \label{numasstri}
\end{eqnarray}
We naturally expect that the magnitudes of the coupling constants $f^{\alpha \beta}$
are between the Yukawa coupling constants $y_{e}$ and $y_{\tau}$ of $e$ and $\tau$, that is
 \begin{eqnarray}
y_{e}\simeq 2.9\times 10^{-6}  \lesssim |f^{\alpha \beta}| \lesssim y_{\tau} \simeq 1.0 \times 10^{-2}.
\label{cupmag}
\end{eqnarray}
The coupling constant $A$ is expected to be as small as the order between 1 eV
and 1 MeV in the models of the large extra dimension at classical level \cite{Ema1, Ema2}.
The masses of the triplet Higgs fields, $M$ is expected to be of the order of the electroweak scale.
When we fix $A$ and $M$ at $A=1$ keV and $M=1$ TeV, for example,
the neutrino masses are of the order, $10^{-5} \sim 10^{-1}$ eV, which can be enough small to
satisfy the results of the neutrino oscillation experiments and WMAP. 
Furthermore, since the lepton number is conserved in
the limit $A \to 0$, the naturalness proposed by 't Hooft guarantees the stability  
of the small $A$ under the radiative corrections.
Thus, we make sure that the smallness of the neutrino is naturally explained in this model.

We finally check how the observations of the $\rho$ parameter restrict
the $SU(2)$ triplet Higgs model. 
The $\rho$ parameter is defined and calculated to be
 \begin{eqnarray}
\rho \equiv \frac{M_{W}^{2}}{M_{Z}^{2}\cos^{2}\theta_{W}}
=\frac{v^{2}+2v_{\Delta}^{2}}{v^{2}+4v_{\Delta}^{2}} \simeq 1-2\frac{v_{\Delta}^{2}}{v^{2}},
\end{eqnarray}
where $\theta_{W}$ is the Weinberg angle and
$M_{W}$ and $M_{Z}$ are the masses of the W and Z gauge bosons.
The constraint on the $\Delta \rho \equiv \rho-1$ by LEP experiments \cite{gri, erl} restricts the ratio of 
VEVs, $v_{\Delta}/v$, as
\begin{eqnarray}
\frac{v_{\Delta}}{v}=\frac{Av}{2\sqrt{2}M^{2}} \ \lesssim \ 0.03. \label{lep}
\end{eqnarray}
It is noted that the constraint (\ref{lep}) gives the upper bounds on the coupling
constant $A$ for the fixed $M$, but does not restrict the coupling constant $f^{\alpha \beta}$ at all.

\section{Washing Out of Baryon Number \label{Washing}}
We want to figure out the condition that the initial baryon number  is not washed out in
the model we consider, assuming that the initial baryon 
or lepton number was generated by some unknown mechanism in the very early universe.
In this section, we obtain the condition in the $SU(2)_{L}$ triplet Higgs model by solving the Boltzmann equation.
In subsection \ref{beq}, we construct the Boltzmann equation  in the $SU(2)_{L}$ triplet
 Higgs model for the case of no CP violation in the lepton sector.  
In subsection \ref{sol}, we solve the Boltzmann equation and derive the time evolution
of the lepton and baryon numbers. In subsection \ref{con}, we obtain the condition 
for the baryon number not to be washed out. 
The way in which the baryon number is washed out in the minimal $SU(5)$ GUT with no CP violation
is analyzed in \cite{Fry1, Fry2}.

\subsection{Boltzmann equation in the $SU(2)_{L}$ triplet Higgs model  \label{beq}}

Since our purpose is to know the time evolution of the difference between the particle number and the anti-particle
number, we define the baryon number $B$, the lepton number $L$, the charged gauge boson number $W$, 
the doublet Higgs number $\Phi$, and the triplet Higgs number $\Delta$ as    
\begin{eqnarray}
B &\equiv& \frac{n_{B}}{s}=\frac{n_{b}-n_{\bar{b}}}{s}=
\sum_{i=u,c,t} \frac{1}{3} \biggl[\frac{n_{u_{iL}}-n_{\bar{u}_{iL}}}{s}+\frac{n_{u_{iR}}-n_{\bar{u}_{iR}}}{s}\biggr]
\nonumber \\
 & & \hspace{30mm} +\sum_{i=d,s,b} \frac{1}{3} \biggl[\frac{n_{d_{iL}}-n_{\bar{d}_{iL}}}{s}
 +\frac{n_{d_{iR}}-n_{\bar{d}_{iR}}}{s}\biggr],  \\
L &\equiv& \frac{n_{L}}{s}=\frac{n_{l}-n_{\bar{l}}}{s}=
\sum_{\alpha=e,\mu,\tau} \biggl[\frac{n_{\nu_{\alpha L}}-n_{\bar{\nu}_{\alpha L}}}{s}
\nonumber \\
& & \hspace{40mm} +\frac{n_{e_{\alpha L}}-n_{\bar{e}_{\alpha L}}}{s}+
\frac{n_{e_{\alpha R}}-n_{\bar{e}_{\alpha R}}}{s}\biggr],  \\
W &\equiv& \frac{n_{W}}{s}=\frac{n_{w^{-}}-n_{w^{+}}}{s},  \\
\Phi &\equiv& \frac{n_{\Phi}}{s}=\frac{n_{\phi^{+}}-n_{\phi^{-}}}{s}+\frac{n_{\phi^{0}}-n_{\phi^{0 \ast}}}{s}, \\
\Delta &\equiv& \frac{n_{\Delta}}{s}=\frac{n_{\xi^{++}}-n_{\xi^{--}}}{s}
+\frac{n_{\xi^{+}}-n_{\xi^{-}}}{s}+\frac{n_{\xi^{0}}-n_{\xi^{0 \ast}}}{s}, 
\end{eqnarray} 
where $n_{X}$ is the number density of a particle $X$ and $s$ is the entropy density in the universe.
We rewrite the  numbers defined above by using the relation described by chemical potential (\ref{che}) as
\begin{eqnarray}
B &\equiv& k \times \biggl[ \sum_{i=u,c,t} (\mu_{u_{iL}}+\mu_{u_{iR}})
                                                  +\sum_{i=d,s,b}(\mu_{d_{iL}}+\mu_{d_{iR}})  \biggr], \\
L &\equiv&  k \times \sum_{\alpha=e,\mu,\tau}
(\mu_{\nu_{\alpha}}+\mu_{\alpha_{L}}+\mu_{\alpha_{R}}),\\
W &\equiv& k \times 4\mu_{w}, \\
\Phi &\equiv& k \times 2(-\mu_{\phi^{-}}+\mu_{\phi^{0}}),  \\
\Delta &\equiv& k \times 2(-\mu_{\xi^{--}}-\mu_{\xi^{-}}+\mu_{\phi^{0}}), 
\end{eqnarray} 
where $\mu_{X}$ is the chemical potential of a particle $X$ and $k$ is defined as $k \equiv T^{2}/3s$.
We next define the total electromagnetic charge $Q$ and the total third component $I_{3}$ of
the $SU(2)_{L}$ isospin as
\begin{eqnarray}
Q &\equiv& \sum_{i=all}Q_{i} \frac{n_{i}-n_{\bar{i}}}{s}=
k \times \biggl[\sum_{i=boson} Q_{i} g_{i}  \mu_{i} +\frac{1}{2}\sum_{i=fermion} 
Q_{i} \mu_{i} g_{i} \biggr]    \nonumber \\
&=& k \times \biggl[ -2\mu_{\phi^{-}}-4\mu_{\xi^{--}}-2\mu_{\xi^{-}}-4\mu_{w^{-}}  
\nonumber \\
& & \hspace{10mm}+2\sum_{i=u,c,t}(\mu_{u_{iL}}+\mu_{u_{iR}})
-\sum_{i=d,s,b}(\mu_{d_{iL}}+\mu_{d_{iR}})
-\sum_{\alpha=e,\mu,\tau}(\mu_{\alpha L}+\mu_{\alpha R})
 \biggr],  \label{qn} \\
I_{3} &\equiv& \sum_{i=all}I_{3}^{i} \frac{n_{i}-n_{\bar{i}}}{s}
= k \times \biggl[ -\mu_{\phi^{-}}-\mu_{\phi^{0}}-2\mu_{\xi^{--}}-2\mu_{\xi^{0}}-4\mu_{w^{-}}  
\nonumber \\
& & \hspace{40mm}+\frac{3}{2} \bigl(\sum_{i=u,c,t}\mu_{u_{L}}-\sum_{i=d,s,b}\mu_{d_{L}} \bigr)
+\frac{1}{2}\sum_{\alpha=e,\mu,\tau}(\mu_{\nu_{\alpha}}-\mu_{\alpha L})
 \biggr],  \label{i3n}
\end{eqnarray} 
where $Q_{i}$ and $I_{3}^{i}$ are the electromagnetic charge and the third component of the isospin 
assigned to the particle $i$, and $g_{i}$ is the internal degree of freedom of the particle.

The time evolution of the particle numbers in a model, of course, is caused by the interactions
 in the model. We enumerate each interaction in the model below.
 The interactions (i) $\sim$ (vi) listed below exist in the standard model.
\begin{itemize} 
\item[(i)] The gauge interaction of the quarks : 
 ${\cal L}=g/\sqrt{2}\bar{u}_{iL}\gamma^{\mu}V_{KM}^{ij}d_{jL}W_{\mu}^{+} +\mbox{h.c.}$ \\
 Here $V_{KM}$ is the Kobayashi-Maskawa matrix. This Lagrangian density causes the processes,
\begin{eqnarray}
 d_{jL} \longleftrightarrow u_{iL} + W_{\mu}^{-}.  \label{(1)}
\end{eqnarray} 

\item[(ii)] The gauge interaction of the leptons : ${\cal L}=g/\sqrt{2} \bar{\nu}_{\alpha L}
\gamma^{\mu}e_{\alpha L}W_{\mu}^{+} +\mbox{h.c.}$ 
\begin{eqnarray}
   \Rightarrow  \ \  e_{\alpha L} \longleftrightarrow \nu_{\alpha L} + W_{\mu}^{-}. \label{(2)}
\end{eqnarray} 

\item[(iii), (iv)] The Yukawa interactions between the up-type-quarks 
(down-type-quarks) and the doublet Higgs fields : ${\cal L}=\hat{y}_{d}^{i} \bar{d}_{iL} d_{iR} 
\phi^{0}+\hat{y}_{u}^{i} \bar{u}_{iL}u_{iR} \phi^{0\ast} +\mbox{h.c.}$
\begin{eqnarray}
   \Rightarrow  \ \  d_{iR} +  \phi^{0} \longleftrightarrow d_{iL},  \ \ \   u_{iR} \longleftrightarrow u_{iL} + 
   \phi^{0}.  \label{(34)}
\end{eqnarray}

\item[(v)] The Yukawa interaction between the leptons and the doublet Higgs fields : \\
${\cal L}=\hat{y}_{e}^{\alpha} \bar{e}_{\alpha L} e_{\alpha R} \phi^{0} +\mbox{h.c.}$
\begin{eqnarray}
   \Rightarrow  \ \  e_{\alpha R} +  \phi^{0} \longleftrightarrow e_{\alpha L}.  \label{(5)}
\end{eqnarray} 

\item[(vi)] The gauge interaction of the doublet Higgs fields : ${\cal L}=\frac{g}{\sqrt{2}} 
(\partial^{\mu} \phi^{-}) \phi^{0} W_{\mu}^{+}+\mbox{h.c.}$
\begin{eqnarray}
   \Rightarrow  \ \  \phi^{-} +\phi^{0} \longleftrightarrow W_{\mu}^{-}.  \label{(6)}
\end{eqnarray} 

The following interactions (vii),(viii),(ix) do not exist in the standard model.
\item[(vii)] The Yukawa interactions between the leptons and the triplet Higgs fields in Eq. (\ref{lld}) :
 \begin{eqnarray}
 &&{\cal L}^{yukawa}_{\nu}= -\frac{1}{2}f^{\alpha \beta}\biggl[ \overline{(\nu^{\alpha})^{c}}
 \nu^{\beta}\xi^{0}-\frac{1}{\sqrt{2}}(\overline{(\nu^{\alpha})^{c}}e^{\beta}+
 \overline{(e^{\alpha})^{c}}\nu^{\beta})\xi^{+}-\overline{(e^{\alpha})^{c}}e^{\beta}\xi^{++}\biggr]
 +\mbox{h.c.}     \nonumber \\
   &\Rightarrow&  \ \xi^{0} \longleftrightarrow \bar{\nu}_{\alpha L} +\bar{\nu}_{\beta L}, \ 
\xi^{+} \longleftrightarrow \bar{\nu}_{\alpha L} +\bar{e}_{\beta L}, \ 
\xi^{++} \longleftrightarrow \bar{e}_{\alpha L} +\bar{e}_{\beta L}.  \label{(7)}
\end{eqnarray}

\item[(viii)] The cubic coupling between the doublet Higgs and the triplet Higgs fields in Eq. (\ref{ppd5}) :
  \begin{eqnarray}
&& {\cal L}^{cubic}=-\frac{1}{2} A \biggl[ -(\phi^{0})^{2}\xi^{0 \ast}-\sqrt{2}
 \phi^{+}\phi^{0}\xi^{-}+(\phi^{+})^{2}\xi^{--}\biggr]
 +\mbox{h.c.}   \nonumber \\
   &\Rightarrow&  \  \xi^{0} \longleftrightarrow \phi^{0} + \phi^{0}, \ 
\xi^{+} \longleftrightarrow \phi^{0}+\phi^{+}, \ 
\xi^{++} \longleftrightarrow \phi^{+}+\phi^{+}.  \label{(8)}
\end{eqnarray}

\item[(ix)] The gauge interaction of the triplet Higgs fields : ${\cal L}=g \bigl[ (\partial^{\mu}
 \xi^{-}) \xi^{0}-(\partial^{\mu} \xi^{--}) \xi^{+} \bigr]W_{\mu}^{+}  +\mbox{h.c.}$
\begin{eqnarray}
\Rightarrow \ \ \xi^{-}+\xi^{0} \longleftrightarrow W_{\mu}^{-},  \ \ \xi^{--} \longleftrightarrow \xi^{-}+W_{\mu}^{-}.
\label{(9)}
\end{eqnarray} 

There also exist the sphaleron-anomaly processes through the non-perturbative effects in the standard model. 
\item[(x)] The sphaleron-anomaly processes
\begin{eqnarray}
u_{iL} u_{iL} d_{iL} \longleftrightarrow \bar{e}_{\alpha L}, \label{sph} \ \ 
u_{iL} d_{iL} d_{iL} \longleftrightarrow \bar{\nu}_{\alpha L}.
\end{eqnarray}

\end{itemize}

We are ready to write down the Boltzmann equations in the $SU(2)_{L}$ triplet Higgs model.
We first  write down the  Boltzmann equation for the time evolution of the lepton number $L$.
The change of the lepton number $L$ is caused by the lepton number violating 
processes, which are the Yukawa interactions between the leptons and the triplet Higgs fields (vii)
 and the sphaleron-anomaly processes (x). We write down the  Boltzmann equation 
for the time evolution of the lepton number density $n_{l}$ in Eq. (\ref{levo})
under the assumption that the phase space density of a particle $X$ is given by the Maxwell-Boltzmann distribution,
\begin{eqnarray}
f(E_{X})=\exp \biggl[-\frac{E_{X}-\mu_{X}}{T} \biggr].
\end{eqnarray}
In  Eq. (\ref{levo}), $M_{f}(f^{\alpha \alpha})$ is the invariant matrix element for the processes,
$\bar{\xi}^{0} \longleftrightarrow \nu_{\alpha L} +\nu_{\alpha L}$ in (\ref{(7)}).
 The other invariant matrix elements, $M_{X}$ are defined in the similar way. 
 Although we can not exactly calculate the invariant matrix element for the sphaleron-anomaly 
 processes $M_{s}$, the obtained results in this section do not depend on the detail of the element
 $M_{s}$, because we require that the sphaleron-anomaly 
 processes are in equilibrium at large temperature region later.
 We can also write down the  Boltzmann equation for the anti-lepton number density $n_{\bar{l}}$
 in the similar way. The difference of the two equations leads to the 
 the Boltzmann equation for the lepton number density $n_{L}(=n_{l}-n_{\bar{l}})$
\begin{eqnarray}
\dot{n}_{L}+3H n_{L}&=& \bigcap \cdot e^{-\frac{E}{T}} \cdot \frac{2}{T}
\nonumber \\
&\biggl[& \sum_{\alpha=e,\mu,\tau} |M_{f}(f^{\alpha \alpha})|^{2} (
-\mu_{\xi^{0}}-2\mu_{\nu_{\alpha}})+2|M_{f}(f^{e \mu})|^{2} (
-\mu_{\xi^{0}}-\mu_{\nu_{e}}-\mu_{\nu_{\mu}})  \nonumber \\
& &+2|M_{f}(f^{e \tau})|^{2} (
-\mu_{\xi^{0}}-\mu_{\nu_{e}}-\mu_{\nu_{\tau}})+2|M_{f}(f^{\mu \tau})|^{2}  (
-\mu_{\xi^{0}}-\mu_{\nu_{\mu}}-\mu_{\nu_{\tau}})\nonumber \\
\nonumber \\
&+&\sum_{\alpha=e,\mu,\tau} |M_{f}(f^{\alpha \alpha})|^{2}
(\mu_{\xi^{-}}-\mu_{\nu_{\alpha}}-\mu_{{\alpha L}})+|M_{f}(f^{e \mu})|^{2} 
(\mu_{\xi^{-}}-\mu_{\nu_{e}}-\mu_{\mu_{L}}) \nonumber \\
& &+|M_{f}(f^{e \tau})|^{2}
(\mu_{\xi^{-}}-\mu_{\nu_{e}}-\mu_{\tau_{L}})+|M_{f}(f^{\mu \tau})|^{2}
 (\mu_{\xi^{-}}-\mu_{\nu_{\mu}}-\mu_{\tau_{L}}) \nonumber \\
& &+|M_{f}(f^{e \mu})|^{2} 
 (\mu_{\xi^{-}}-\mu_{e_{L}}-\mu_{\nu_{\mu}})+|M_{f}(f^{e \tau})|^{2}
 (\mu_{\xi^{-}}-\mu_{e_{L}}-\mu_{\nu_{\tau}})  \nonumber \\
& &+|M_{f}(f^{\mu \tau})|^{2} 
 (\mu_{\xi^{-}}-\mu_{\mu_{L}}-\mu_{\nu_{\tau}})  \nonumber \\
\nonumber \\
&+&\sum_{\alpha=e,\mu,\tau} |M_{f}(f^{\alpha \alpha})|^{2} 
(\mu_{\xi^{--}}-2\mu_{\alpha_{L}})+2|M_{f}(f^{e \mu})|^{2} 
(\mu_{\xi^{--}}-\mu_{e_{L}}-\mu_{\mu_{L}})  \nonumber \\
& &+2|M_{f}(f^{e \tau})|^{2}
(\mu_{\xi^{--}}-\mu_{e_{L}}-\mu_{\tau_{L}})+2|M_{f}(f^{\mu \tau})|^{2} 
(\mu_{\xi^{--}}-\mu_{\mu_{L}}-\mu_{\tau_{L}})  \nonumber \\
\nonumber \\
&+& \int d \Pi_{3} \frac{\delta^{4}(p_{X}-p_{1}-p_{2}-p_{3})}{\delta^{4}(p_{X}-p_{1}-p_{2})}  \nonumber \\
& &\sum_{\alpha=e,\mu,\tau} \frac{-1}{2} |M_{s}|^{2} 
(2\mu_{u_{L}}+\mu_{d_{L}}+\mu_{\alpha_{L}}+\mu_{u_{L}}+2\mu_{d_{L}}+\mu_{\nu_{\alpha}})
\biggr],  \label{Levo}
\end{eqnarray}
where $\dot{n}_{L} \equiv dn_{L}(t)/dt$, 
$\bigcap \equiv \int d \Pi_{X} d {\Pi}_{1} d \Pi_{2} (2 \pi)^{4} \delta^{4}(p_{X}-p_{1}-p_{2})$,
$\int d \Pi_{i}\equiv \int \frac{d^{3}p_{i}}{(2\pi)^{3}}\frac{1}{2E_{i}}$, and $H$ is the Hubble 
parameter. Here we use the following relations,  
\begin{eqnarray}
& & f(X)-f(\bar{X})=\exp\biggl[-\frac{E-\mu_{X}}{T}\biggr]-\exp \biggl[-\frac{E+\mu_{X}}{T} \biggr]
\simeq e^{-\frac{E}{T}} \cdot \frac{2\mu_{X}}{T},  \\
&&f(1)f(2)-f(\bar{1})f(\bar{2})\simeq e^{-\frac{E}{T}} \cdot \frac{2(\mu_{1}+\mu_{2})}{T},  \\
&&f(1)f(2)f(3)-f(\bar{1})f(\bar{2})f(\bar{3})
\simeq e^{-\frac{E}{T}} \cdot \frac{2(\mu_{1}+\mu_{2}+\mu_{3})}{T},
\end{eqnarray}
where the approximation $\mu_{i}/T \ll 1$ is used.

We next write down the Boltzmann equation for the baryon number $B$.
The change of the baryon number $B$ is caused by the baryon number violating 
process, which is the sphaleron-anomaly processes (x) only in this model.
We write down the  Boltzmann equation 
for the baryon number density $n_{b}$ in Eq. (\ref{bevo}).
 We can also write down the  Boltzmann equation for the anti-baryon number density $n_{\bar{b}}$.
 The difference of the two equations leads to
 the  Boltzmann equation for the baryon number density $n_{B}(=n_{b}-n_{\bar{b}})$
\begin{eqnarray}
\dot{n}_{B}+3H n_{B}&=&\int d \Pi_{X} d {\Pi}_{1} d \Pi_{2} d \Pi_{3}
(2 \pi)^{4} \delta^{4}(p_{X}-p_{1}-p_{2}-p_{3}) \cdot e^{-\frac{E}{T}} \cdot \frac{2}{T}
\nonumber \\
& &\sum_{\alpha=e,\mu,\tau} \frac{-1}{2} |M_{s}|^{2} 
(\mu_{\alpha_{L}}+2\mu_{u_{L}}+\mu_{d_{L}}+\mu_{\nu_{\alpha}}+\mu_{u_{L}}+2\mu_{d_{L}}).
\label{Bevo}
\end{eqnarray}

We write down the  Boltzmann equation for the triplet Higgs number $\Delta$.
The change of $\Delta$ is caused by the interactions (vii) and (viii).
We write down the  Boltzmann equation  for the triplet Higgs number density $n_{\delta}$ 
in Eq. (\ref{devo}). We obtain the  Boltzmann equation for $n_{\Delta}(=n_{\delta}-n_{\bar{\delta}})$
\begin{eqnarray}
\dot{n}_{\Delta}+3H n_{\Delta}&=&  \bigcap \cdot e^{-\frac{E}{T}} \cdot \frac{2}{T}
\nonumber \\
&\biggl[& \sum_{\alpha=e,\mu,\tau} \frac{1}{2}|M_{f}(f^{\alpha \alpha})|^{2} 
(-2 \mu_{\nu_{\alpha}}-\mu_{\xi^{0}})+|M_{f}(f^{e \mu})|^{2} 
(-\mu_{\nu_{e}}-\mu_{\nu_{\mu}}-\mu_{\xi^{0}})  \nonumber \\
& &+|M_{f}(f^{e \tau})|^{2} 
(-\mu_{\nu_{e}}-\mu_{\nu_{\tau}}-\mu_{\xi^{0}})+|M_{f}(f^{\mu \tau})|^{2}
(-\mu_{\nu_{\mu}}-\mu_{\nu_{\tau}}-\mu_{\xi^{0}})   \nonumber \\
\nonumber \\
&+&\sum_{\alpha=e,\mu,\tau}\frac{1}{2} |M_{f}(f^{\alpha \alpha})|^{2}
(-\mu_{\nu_{\alpha}}-\mu_{\alpha L}+\mu_{\xi^{-}}) \nonumber \\
& &+\frac{1}{2}|M_{f}(f^{e \mu})|^{2} 
(-\mu_{\nu_{e}}-\mu_{\mu L}+\mu_{\xi^{-}})+\frac{1}{2}|M_{f}(f^{e \tau})|^{2} 
(-\mu_{\nu_{e}}-\mu_{\tau L}+\mu_{\xi^{-}})  \nonumber \\
& &+\frac{1}{2}|M_{f}(f^{\mu \tau})|^{2} 
(-\mu_{\nu_{\mu}}-\mu_{\tau L}+\mu_{\xi^{-}})+\frac{1}{2}|M_{f}(f^{e \mu})|^{2} 
(-\mu_{e L}-\mu_{\nu_{\mu}}+\mu_{\xi^{-}})  \nonumber \\
& &+\frac{1}{2}|M_{f}(f^{e \tau})|^{2} 
(-\mu_{e L}-\mu_{\nu_{\tau}}+\mu_{\xi^{-}})+\frac{1}{2}|M_{f}(f^{\mu \tau})|^{2}
 (-\mu_{\mu L}-\mu_{\nu_{\tau}}+\mu_{\xi^{-}})   \nonumber \\
\nonumber \\
&+&\sum_{\alpha=e,\mu,\tau}\frac{1}{2} |M_{f}(f^{\alpha \alpha})|^{2} 
(-2\mu_{\alpha L}+\mu_{\xi^{--}})+|M_{f}(f^{e \mu})|^{2} 
(-\mu_{e L}-\mu_{\mu L}+\mu_{\xi^{--}})  \nonumber \\
& &+|M_{f}(f^{e \tau})|^{2}
(-\mu_{e L}-\mu_{\tau L}+\mu_{\xi^{--}})+|M_{f}(f^{\mu \tau})|^{2} 
(-\mu_{\mu L}-\mu_{\tau L}+\mu_{\xi^{--}})   \nonumber \\
\nonumber \\
&+&\frac{1}{2}|M_{A}(A)|^{2} 
(2\mu_{\phi^{0}}-\mu_{\xi^{0}})+\frac{1}{2}|M_{A}(A)|^{2} 
(\mu_{\phi^{0}}-\mu_{\phi^{-}}+\mu_{\xi^{-}})  \nonumber \\
& &+\frac{1}{2}|M_{A}(A)|^{2} 
(-2\mu_{\phi^{-}}+\mu_{\xi^{--}})\biggr].  \label{Devo}
\end{eqnarray}

We similarly write down the  Boltzmann equation for the doublet Higgs number $\Phi$.
The change of $\Phi$ is caused by the interactions (iii), (iv), (v), and (viii).
The charged current processes in companion with (iii), (iv), (v) also contribute to the changes of the number $\Phi$.
We write down the  Boltzmann equation  for the doublet Higgs number density $n_{\phi}$ 
in Eq. (\ref{pevo}). The  Boltzmann equation for $n_{\Phi}(=n_{\phi}-n_{\bar{\phi}})$ is
\begin{eqnarray}
\dot{n}_{\Phi}+3H n_{\Phi}&=&  \bigcap  \cdot e^{-\frac{E}{T}} \cdot \frac{2}{T}
\nonumber \\
& \biggl[& \sum_{i=d,s,b} |M_{d}(y_{d}^{i})|^{2}
(-\mu_{d_{iR}}+\mu_{d_{iL}}-\mu_{\phi^{0}})   \nonumber \\
& & +\sum_{i=u,c,t} |M_{u}(y_{u}^{i})|^{2}
(\mu_{u_{iR}}-\mu_{u_{iL}}-\mu_{\phi^{0}})   \nonumber \\
 & & +\sum_{\alpha=e,\mu,\tau} |M_{e}(y_{e}^{\alpha})|^{2}
(-\mu_{\alpha R}+\mu_{\alpha L}-\mu_{\phi^{0}})  \nonumber \\
 \nonumber\\
 & & +\sum_{i} |M_{d}(y_{d}^{i})|^{2}
 (-\mu_{d_{iR}}+\mu_{u_{iL}}+\mu_{\phi^{-}})   \nonumber \\
 & & +\sum_{i} |M_{u}(y_{u}^{i})|^{2}
  (\mu_{u_{iR}}-\mu_{d_{iL}}+\mu_{\phi^{-}})  \nonumber \\
& & +\sum_{\alpha=e,\mu,\tau} |M_{e}(y_{e}^{\alpha})|^{2}
(-\mu_{\alpha R}+\mu_{\nu_{\alpha}}+\mu_{\phi^{-}})  \nonumber \\
 \nonumber\\
 & &+|M_{A}(A)|^{2} [(-\mu_{\xi^{--}}+2\mu_{\phi^{-}})
+(-\mu_{\xi^{-}}-\mu_{\phi^{0}}+\mu_{\phi^{-}}) \nonumber\\
& &\hspace{65mm}+(\mu_{\xi^{0}}-2\mu_{\phi^{0}})] \biggr].   \label{Pevo}
\end{eqnarray}

We finally write down the  Boltzmann equation for the number of the charged gauge boson $W$.
The change of $W$ is caused by the interactions (i), (ii), (vi), and (ix).
We write down the  Boltzmann equation  for the number density $n_{w^{-}}$ in Eq. (\ref{wevo}). 
The obtained Boltzmann equation for $n_{W}(=n_{w^{-}}-n_{w^{+}})$ reads as
\begin{eqnarray}
\dot{n}_{W}+3H n_{W}&=&  \bigcap  \cdot e^{-\frac{E}{T}} \cdot \frac{2}{T}
\nonumber \\
& \biggl[& \sum_{i,j} |M_{g}(\frac{g}{\sqrt{2}}V_{KM}^{ij})|^{2}
(-\mu_{u_{iL}}+\mu_{d_{jL}}-\mu_{w^{-}})  \nonumber \\
 & & +\sum_{\alpha=e,\mu.\tau} |M_{g}(\frac{g}{\sqrt{2}})|^{2}
 (-\mu_{\nu_{\alpha}}+\mu_{\alpha_{L}}-\mu_{w^{-}})  \nonumber \\
 & & +|M_{g}(\frac{g}{\sqrt{2}}k_{\mu})|^{2}
(\mu_{\phi^{-}}+\mu_{\phi^{0}}-\mu_{w^{-}})  \nonumber \\
 & &  +|M_{g}({g}k_{\mu})|^{2}
(\mu_{\xi^{-}}+\mu_{\xi^{0}}-\mu_{w^{-}})  \nonumber \\
& &  +|M_{g}({g}k_{\mu})|^{2}
(\mu_{\xi^{--}}-\mu_{\xi^{-}}-\mu_{w^{-}})\biggr].      \label{Wevo}
\end{eqnarray}

 Since the $SU(2)_{L}$ triplet Higgs model is within the framework of the $SU(2)_{L} \otimes U(1)_{Y}$
 gauge theory, the total electromagnetic charge $Q$ and the total third component of the $SU(2)_{L}$
 isospin $I_{3}$ are conserved. It is natural to assume that the total quantity of the conserved charge
 associated with the gauge symmetry vanish. Under this assumption Eqs. (\ref{qn}) and  (\ref{i3n}) reduce to the 
 following relations,
\begin{eqnarray}
Q=0 \Leftrightarrow && -2\mu_{\phi^{-}}-4\mu_{\xi^{--}}-2\mu_{\xi^{-}}-4\mu_{w^{-}}
+2\sum_{i=u,c,t}(\mu_{u_{i L}}+\mu_{u_{i R}})
\nonumber \\
& & \hspace{20mm} -\sum_{i=d,s,b}
(\mu_{d_{i L}}+\mu_{d_{i R}})-\sum_{\alpha=e,\mu,\tau}(\mu_{\alpha L}+\mu_{\alpha R})=0
\label{qn0}
\end{eqnarray}    
and
\begin{eqnarray}
I_{3}=0 \Leftrightarrow & &  -\mu_{\phi^{-}}-\mu_{\phi^{0}}-2\mu_{\xi^{--}}-2\mu_{\xi^{0}}-4\mu_{w^{-}}  
\nonumber \\
& & \hspace{2mm}+\frac{3}{2} \biggl(\sum_{i=u,c,t} \mu_{u_{i L}}-\sum_{i=d,s,b}\mu_{d_{i L}} \biggr)
+\frac{1}{2}\sum_{\alpha=e,\mu,\tau}(\mu_{\nu_{\alpha}}-\mu_{\alpha L})=0.  \label{3n0}
\end{eqnarray}

We can regard the interactions (i) $\sim$ (vi) and (ix) as being in enough equilibrium.
Since the creation and annihilation rates of a particle in in-equilibrium processes
are equal, the in-equilibrium processes do not change the number of a particle.
This observation yields the following relations about the chemical potentials from (\ref{(1)}), (\ref{(2)}),
 (\ref{(34)}), (\ref{(5)}), (\ref{(6)}), and (\ref{(9)}),
\begin{eqnarray}
\mu_{d_{L}}&=&\mu_{u_{L}}+\mu_{w^{-}}, \  \ \ \mu_{\alpha_{L}}=\mu_{\nu_{\alpha}}+\mu_{w^{-}},  
\ \ \ \ \ \ \ \mu_{d_{R}}=\mu_{u_{L}}+\mu_{w^{-}} -\mu_{\phi^{0}}, \nonumber \\ 
\mu_{u_{R}}&=&\mu_{u_{L}}+\mu_{\phi^{0}},   \ 
 \ \ \mu_{\alpha_R}=\mu_{\nu_{\alpha}}+\mu_{w^{-}}-\mu_{\phi^{0}},
\ \ \ \mu_{\phi^{-}}=\mu_{w^{-}}-\mu_{\phi^{0}}, \label{eqc}  \\ 
\ \mu_{\xi^{-}}&=&\mu_{w^{-}}-\mu_{\xi^{0}}, 
\ \ \ \mu_{\xi^{--}}=2\mu_{w^{-}}-\mu_{\xi^{0}},    \nonumber 
\end{eqnarray}
where we take the seven parameters, 
$(\mu_{w^{-}}, \mu_{u_{L}}, \mu_{\nu_{\alpha}}, \mu_{\phi^{0}}, \mu_{\xi^{0}} )$
as the independent parameters out of the nineteen ones.
In Eq. (\ref{eqc}), since the in-equilibrium processes (\ref{(1)}) mix the up-type flavors of
the left-handed quarks enough, the chemical potentials of them become all equal :
$\mu_{u_{i L}}\equiv \mu_{u_{L}} \ (i=u, c, t)$.
Furthermore, since the mixed flavors are transmitted to the up-type flavors of
the right-handed quarks through the interactions (\ref{(34)}),
 the chemical potentials of them also become identical :
 $\mu_{u_{i R}}\equiv \mu_{u_{R}} \ (i=u, c, t)$.
The similar processes lead to the similar relations for the chemical potentials of the down-type 
quarks as $\mu_{d_{i L}}\equiv \mu_{d_{L}}$ and $\mu_{d_{i R}}\equiv \mu_{d_{R}} \ (i=d, s, b)$.
Eq. (\ref{3n0}) restricted by Eqs. (\ref{eqc}) is transformed to the relation,
\begin{eqnarray}
\mu_{w^{-}}=0.  \label{w0}
\end{eqnarray}
Inversely, Eqs. (\ref{eqc}) restricted by Eq. (\ref{w0}) are written as,
\begin{eqnarray}
\mu_{d_{L}}&=&\mu_{u_{L}},  \hspace{15mm} \mu_{\alpha_{L}}=\mu_{\nu_{\alpha}}, \ \ \ 
\hspace{11mm} \mu_{d_{R}}=\mu_{u_{L}} -\mu_{\phi^{0}}, \nonumber\\
\mu_{u_{R}}&=&\mu_{u_{L}}+\mu_{\phi^{0}},\ \ \ 
\mu_{\alpha_R}=\mu_{\nu_{\alpha}}-\mu_{\phi^{0}}, \ \ \ 
\mu_{\phi^{-}}=-\mu_{\phi^{0}},  \label{eqcem}  \\
\mu_{\xi^{-}}&=&-\mu_{\xi^{0}}, \hspace{11mm} \mu_{\xi^{--}}=-\mu_{\xi^{0}}, \nonumber 
\end{eqnarray}
where the six parameters $(\mu_{u_{L}}, \mu_{\nu_{\alpha}}, \mu_{\phi^{0}}, \mu_{\xi^{0}})$ are
taken to be the independent ones. Eq. (\ref{qn0}) can be written by the  six 
independent parameters as
\begin{eqnarray}
7\mu_{\phi^{0}}+3\mu_{\xi^{0}}+6\mu_{u_{L}}-2\mu_{\nu}=0,  \label{q01}
\end{eqnarray}
where we define $\mu_{\nu}$ as
\begin{eqnarray}
\mu_{\nu} \equiv \sum_{\alpha=e,\mu.\tau} \mu_{\nu_{\alpha}}.
\end{eqnarray}
Eq. (\ref{q01}) makes the number of the independent parameters reduce from six to five.
The right hand sides of the Boltzmann equations (\ref{Levo}), (\ref{Bevo}), (\ref{Devo}), (\ref{Pevo}), and (\ref{Wevo})
can be rewritten by the six parameters  as in Eq. (\ref{Leqb}), (\ref{Beqb}), (\ref{Deqb}), (\ref{Peqb}), and (\ref{Weqb}),
where we use the relation,
\begin{eqnarray}
s\frac{dL(t)}{dt}=\frac{dn_{L}(t)}{dt}+3n_{L}(t)H.
\end{eqnarray}
For simplicity, we further assume that the three lepton flavors, $e, \mu, \tau$ are
identical, that is,
\begin{eqnarray}
f^{ee}=f^{\mu \mu}=f^{\tau \tau}=f^{e \mu}=f^{e\tau}=f^{\mu \tau} \equiv f, \nonumber \\
\mbox{and}
\ \ \mu_{\nu_{e}}=\mu_{\nu_{\mu}}=\mu_{\nu_{\tau}}  \Rightarrow  \ \mu_{\nu}=3\mu_{\nu_{\alpha}}.  
\label{emt}
\end{eqnarray} 
Since Eqs. (\ref{emt}) makes the number of the independent parameters reduce from five to three,
we can take  ($L, B, \Delta$) to be the three independent parameters : 
 \begin{eqnarray}
\mu_{\phi^{0}}&=&\frac{1}{k} \cdot \frac{1}{4}\Phi=\frac{1}{k}
\biggl[-\frac{1}{24}B+\frac{1}{18}L-\frac{1}{12}\Delta \biggr],  \label{mup} \\
\mu_{\xi^{0}}&=&\frac{1}{k} \cdot \frac{1}{6}\Delta,  \label{mux} \\
\mu_{u_{L}}&=&\frac{1}{k} \cdot \frac{1}{12}B,  \\
\mu_{\nu}&=&\frac{1}{k}\biggl[-\frac{1}{24}B+\frac{7}{18}L-\frac{1}{12}\Delta \biggr]. 
\label{munu}
\end{eqnarray} 
The Boltzmann equations (\ref{Leqb}), (\ref{Beqb}), (\ref{Deqb}), and (\ref{Peqb})
can be also rewritten by the three parameters ($L, B, \Delta$) as
\begin{eqnarray}
s\frac{d}{dt}\left(\begin{array}{c}
                                   L \\
                                    B  \\
                                     \Delta 						 
									 \end{array}\right)=\bigcap \frac{2 \cdot e^{-\frac{E}{T}}}{T\cdot k}
\left(\begin{array}{ccc}
                               -7m_{f}-\frac{7}{9}m_{s}       & \frac{3}{4}m_{f} -\frac{17}{12}m_{s} & -3m_{f}+\frac{1}{6}m_{s} \\
                                -\frac{7}{9}m_{s}     &  -\frac{17}{12}m_{s}   & \frac{1}{6}m_{s} \\
                                    -\frac{7}{2}m_{f}+\frac{1}{6}m_{A} & \frac{3}{8}m_{f} -\frac{1}{8}m_{A} & -\frac{3}{2}m_{f}
									-\frac{1}{2}m_{A}					 
									 \end{array}\right)
									 \left(\begin{array}{c}
                                   L \\
                                    B  \\
                                     \Delta 						 
									 \end{array}\right),       \nonumber \\   \label{bol3}
\end{eqnarray} 
where
\begin{eqnarray}
m_{f} &\equiv& |M_{f}(f)|^{2},  \\
m_{A} &\equiv& |M_{A}(A)|^{2},  \\
m_{s} &\equiv&  \int d \Pi_{3} \frac{\delta^{4}(p_{X}-p_{1}-p_{2}-p_{3})}{\delta^{4}(p_{X}-p_{1}-p_{2})} 
\frac{1}{2}|M_{s}|^{2}.
\end{eqnarray} 
Let us note that the determinant of the matrix in Eq. (\ref{bol3}),
\begin{eqnarray}
 \begin{array}{|ccc|}
                         -7m_{f}-\frac{7}{9}m_{s}       & \frac{3}{4}m_{f} -\frac{17}{12}m_{s} & -3m_{f}+\frac{1}{6}m_{s} \\
                          -\frac{7}{9}m_{s}     &  -\frac{17}{12}m_{s}   & \frac{1}{6}m_{s} \\
                           -\frac{7}{2}m_{f}+\frac{1}{6}m_{A} & \frac{3}{8}m_{f} -\frac{1}{8}m_{A} & -\frac{3}{2}m_{f}
					-\frac{1}{2}m_{A}
					\end{array}=-\frac{51}{8}m_{f} \cdot m_{A} \cdot m_{s}.
\end{eqnarray} 
The in-equilibrium sphaleron processes (\ref{sph}) at the temperature region 
between 100 GeV and $10^{12}$ GeV lead to
\begin{eqnarray}
9\mu_{u_{L}}+\mu_{\nu}=0,   \label{erspha}
\end{eqnarray} 
which makes the number of the independent parameters reduce from three to two.
We take  ($L, \Delta$) to be the two independent parameters and can reduce
the Boltzmann equation (\ref{bol3}) to 
\begin{eqnarray}
s\frac{d}{dt} \left( \begin{array}{c}
L \\
\Delta 
\end{array} \right)=\bigcap \frac{2 \cdot e^{-\frac{E}{T}}}{T\cdot k}
\frac{3}{17} \left( \begin{array}{cc}
  -42m_{f}   & -\frac{33}{2}m_{f}  \\
  -21m_{f}+\frac{4}{3}m_{A} & -\frac{33}{4}m_{f} -\frac{35}{12}m_{A}
  \end{array} \right)
  \left( \begin{array}{c} L \\ \Delta 
  \end{array} \right) .   \label{bol2}
\end{eqnarray} 
We rewrite the Boltzmann equation (\ref{bol2}) into a simple form, 
\begin{eqnarray}
s\frac{d}{dt}\vec{N}=I M_{2} \vec{N},   \label{bol22}
\end{eqnarray} 
where we define $\vec{N}, I$ and $M_{2}$ as
\begin{eqnarray}
\vec{N} &\equiv& \left(\begin{array}{c}
 L \\
 \Delta
 \end{array}\right),   \ \ I \equiv \bigcap
 \frac{2 \cdot e^{-\frac{E}{T}}}{T\cdot k}, \\
 \mbox{and} \ \ 
 M_{2} &\equiv& \frac{3}{17}\left(\begin{array}{cc}
  -42m_{f}     & -\frac{33}{2}m_{f}  \\
 -21m_{f}+\frac{4}{3}m_{A} & -\frac{33}{4}m_{f} -\frac{35}{12}m_{A} 
 \end{array}\right).
\end{eqnarray} 
Note that the determinant of the matrix, $ M_{2}$, is 
\begin{eqnarray}
det(M_{2})=\biggl(\frac{3}{17} \biggr)^{2} \cdot \frac{289}{2} \cdot m_{f} \cdot m_{A}. 
\end{eqnarray} 

 Here, the baryon number is also written by the linear combination of the two independent parameters, $L$ and $\Delta$
as
\begin{eqnarray}
B=-\frac{28}{51}L+\frac{2}{17}\Delta.   \label{bnum}
\end{eqnarray}

\subsection{Solution of the Boltzmann equation  \label{sol}}
In this subsection, we solve the Boltzmann equation (\ref{bol22}). To solve Eq. (\ref{bol22}), 
we need to know the $IM_{2}$ as a function of the time $t$ or the temperature $T$.
The factor $I\cdot m_{f}$ is rewritten as
\begin{eqnarray}
I \cdot m_{f} =\frac{6s}{T^{3}} \int \frac{d^{3}p_{\xi}}{(2\pi)^{3}} \Gamma (\xi^{0} \to \bar{\nu}_{e} \bar{\nu}_{\mu})
 e^{-\frac{E}{T}}, 
\end{eqnarray} 
where the decay rate $\Gamma (\xi^{0} \to \bar{\nu}_{e} \bar{\nu}_{\mu})$ is defined as 
\begin{eqnarray}
\Gamma (\xi^{0} \to \bar{\nu}_{e} \bar{\nu}_{\mu})\equiv
\frac{1}{2E}\int  \frac{d^{3}p_{1}}{(2\pi)^{3}}\frac{1}{2E_{1}}  \frac{d^{3}p_{2}}{(2\pi)^{3}}\frac{1}{2E_{2}} 
(2 \pi)^{4} \delta^{4}(p_{X}-p_{1}-p_{2})  |M_{f}(f)|^{2}
\end{eqnarray}
and the triplet Higgs field $\xi_{0}$ has the energy $E=\sqrt{|\vec{p}_{\xi}|^{2}+M^{2}}$.
We define the thermal averaged decay rate of the decay process, $\xi^{0} \to \bar{\nu}_{e} \bar{\nu}_{\mu}$
as 
\begin{eqnarray}
 \Gamma_{f} \equiv \langle \Gamma (\xi^{0} \to \bar{\nu}_{e} \bar{\nu}_{\mu}) \rangle =
\frac{\int \frac{d^{3}p_{\xi}}{(2\pi)^{3}} \Gamma (\xi^{0} \to \bar{\nu}_{e} \bar{\nu}_{\mu})
 e^{-\frac{E}{T}} }{\int \frac{d^{3}p_{\xi}}{(2\pi)^{3}}e^{-\frac{E}{T}}} 
\end{eqnarray}
and rewrite the factor $I\cdot m_{f}$ as
\begin{eqnarray}
 I \cdot m_{f}=\frac{6s}{T^{3}} \cdot n_{M}^{eq}(T) \cdot \Gamma_{f},  
\end{eqnarray}
where 
\begin{eqnarray}
n_{M}^{eq}(T) \equiv\int \frac{d^{3}p_{\xi}}{(2\pi)^{3}}e^{-\frac{E}{T}}. 
\end{eqnarray}
Similarly, we obtain the factor $I\cdot m_{A}$ as
\begin{eqnarray}
 I \cdot m_{A}=\frac{6s}{T^{3}} \cdot n_{M}^{eq}(T) \cdot \Gamma_{A}, 
\end{eqnarray}
where $\Gamma_{A}$ is the thermal averaged decay rate : 
\begin{eqnarray}
 \Gamma_{A} \equiv 2 \langle \Gamma (\xi^{0} \to \phi^{0} \phi^{0}) \rangle. 
\end{eqnarray}
Using the thermal averaged decay rates $\Gamma_{f}$ and $\Gamma_{A}$,
we rewrite Eq. (\ref{bol22}) as  
\begin{eqnarray}
\frac{d}{dt}\left(\begin{array}{c}
L \\
 \Delta 
 \end{array}\right)= \frac{n_{M}^{eq}}{T^{3}} \cdot \frac{18}{17}\left(\begin{array}{cc}
 -42 \Gamma_{f}     & -\frac{33}{2} \Gamma_{f}  \\
 -21\Gamma_{f}+\frac{4}{3} \Gamma_{A} & -\frac{33}{4} \Gamma_{f} -\frac{35}{12}
 \Gamma_{A} \end{array}\right)
 \left(\begin{array}{c}
 L \\
 \Delta 
 \end{array}\right).  \label{bolld}
\end{eqnarray}

To solve the Eq. (\ref{bolld}) we need to know the temperature dependence of the 
$\Gamma_{f}$ and $\Gamma_{A}$.
The decay rate $\Gamma (\xi^{0} \to \bar{\nu}_{e} \bar{\nu}_{\mu})$ is 
exactly calculated as
\begin{eqnarray}
\Gamma (\xi^{0} \to \bar{\nu}_{e} \bar{\nu}_{\mu})=\frac{1}{2\pi}|f^{e \mu}|^{2}
\frac{M^{2}}{E} \equiv \frac{\alpha}{E}. \label{alpha3}
\end{eqnarray}

We estimate $\Gamma_{f}$ in the two cases, (i)$T > M$ and (ii)$ T < M$.
 \begin{itemize}
 \item[(i)] $T > M$  \par
We do relativistic approximation, $M \simeq 0$, which leads to
\begin{eqnarray}
\Gamma_{f}=\frac{\int \frac{d^{3}p_{\xi}}{(2\pi)^{3}} \Gamma (\xi^{0} \to \bar{\nu}_{e}
 \bar{\nu}_{e})e^{-\frac{E}{T}}}{n_{M}^{eq}(T)}=\frac{\alpha T^{2}/ 2\pi^{2} }{T^{3}/ \pi^{2}}
 =\frac{\alpha}{2 T}.  
 \end{eqnarray} 
\item[(ii)] $T < M$ \par
We do non-relativistic approximation, $E \simeq M+p^{2}/2M$ and obtain
\begin{eqnarray}
\Gamma_{f}=\frac{\int \frac{d^{3}p}{(2\pi)^3} \Gamma (\xi^{0} \to \bar{\nu}_{e} \bar{\nu}_{e}) 
e^{-\frac{E}{T}}}{n_{M}^{eq}(T) }=\frac{\alpha n_{M}^{eq}/M}
{n_{M}^{eq}}=\frac{\alpha}{M}.  
\end{eqnarray}
\end{itemize}

The decay rate $\Gamma (\xi^{0} \to \phi^{0} \phi^{0} )$ is 
exactly calculated as
\begin{eqnarray}
2 \Gamma (\xi^{0} \to \phi^{0} \phi^{0} )=\frac{1}{8\pi EM}\sqrt{\frac{M^{2}}{4}-m_{\phi^{0}}^{2}}
 \ |A|^{2}\simeq \frac{1}{16\pi}\frac{|A|^{2}}{E} \equiv \frac{\beta}{E}.   \label{beta3}
\end{eqnarray}
We similarly estimate $\Gamma_{A}$ in the two cases as
\begin{eqnarray}
\Gamma_{A}=\left\{ \begin{array}{ll}
 \beta/2T  & (T>M),  \\
\beta/M  & (T<M).
 \end{array}  \right.
 \label{phidecaytri}
\end{eqnarray}

Since the temperature dependence of the right hand side of Eq. (\ref{bolld}) is revealed,
we rewrite the left hand side of the equation as a function of  temperature T using the relation,
\begin{eqnarray}
\frac{d}{dt}=-qT^{3}\frac{d}{dT},
\end{eqnarray}
where we use the relation,
\begin{eqnarray}
H=qT^{2}=\frac{1}{2t}.  
\end{eqnarray}
Here the constant q is defined as
\begin{eqnarray}
q \equiv 1.66 \sqrt{g_{\ast}} \frac{1}{M_{Pl}},   \label{qvalu}
\end{eqnarray}
with $M_{Pl}$ being the Plank mass and $g_{\ast}$ is the total degrees of freedom of effectively 
massless particles. For the following analysis, we introduce $K_{f}$ and $K_{A}$ as
\begin{eqnarray}
K_{f} \equiv  \frac{\Gamma_{f}}{H} \bigg|_{T=M}=\frac{\alpha}{2qM^{3}} \ \ \ 
\mbox{and} \ \ \  K_{A} \equiv  \frac{\Gamma_{A}}{H} \bigg|_{T=M}=\frac{\beta}{2qM^{3}}.
\label{orcon}
\end{eqnarray}
Using the constants $K_{f}$ and $K_{A}$, 
we transform Eq. (\ref{bolld}) into the Boltzmann equation which has the only one parameter T,
\begin{eqnarray}
\frac{d}{dT}\vec{N} = f(T) \cdot M \vec{N},    \label{bolf}
\end{eqnarray} 
where the matrix $M$ and the function $f(T)$ are defined as
\begin{eqnarray}
M&\equiv&\left(\begin{array}{cc}
                               2 K_{f}     & \frac{11}{14} K_{f}  \\
                                    K_{f}-\frac{4}{63}K_{A} & \frac{11}{28} K_{f} +\frac{5}{36} K_{A} 			 
									 \end{array}\right)
									 \equiv \left(\begin{array}{cc}
                            a & d\\
                                   c & b			 
									 \end{array}\right)  \\
									 \mbox{and} \ \ 
f(T) &\equiv& \left\{ \begin{array}{ll}
                                    \gamma \frac{M^{3}}{T^{4}}   & (T>M) \\
                                     \delta \frac{M^{7/2}}{T^{9/2}}e^{-\frac{M}{T}} & (T<M) 						 
									 \end{array}\right.    \label{ft}
\end{eqnarray} 
The constants $\gamma$ and $\delta$ in Eq. (\ref{ft}) are defined as
\begin{eqnarray}
\gamma=\frac{378}{17} \cdot \frac{1}{\pi^{2}} \simeq 2.26
\ \ \ \mbox{and} \ \ \ \delta=\frac{378}{17}\cdot \frac{1}{\pi^{3/2}\cdot \sqrt{2}} \simeq 2.83. 
\end{eqnarray}

We can solve Eq. (\ref{bolf}) by diagonalizing the matrix $M$.
The equation of the eigenvalue determines the two eigenvalues as
\begin{eqnarray}
|M-\lambda I|=0  \ \   \Leftrightarrow  \ \ \lambda =\frac{a+b\pm D}{2},
\end{eqnarray} 
where
\begin{eqnarray}
D \equiv \sqrt{(a-b)^{2}+4cd}.
\end{eqnarray} 
Using the transformation, 
\begin{eqnarray}
\vec{N}^{'}\equiv  \left(\begin{array}{c}
                                   X \\
                                    Y 						 
									 \end{array}\right)=V^{-1} \vec{N}, \ \ \ \mbox{with} \ \ 
									 V \equiv \left(\begin{array}{cc}
                              2d   &  -2d \\
                                D+b-a &	D+a-b
									 \end{array}\right), \label{eigenxy}
\end{eqnarray} 
Eq. (\ref{bolf}) is transformed into the following one,
\begin{eqnarray}
\frac{d}{dT} \vec{N}^{'} =f(T)\cdot \hat{M}  \vec{N}^{'},   \label{bold}
\end{eqnarray} 
where the matrix $\hat{M}$ is diagonalized as 
\begin{eqnarray}
\hat{M}=V^{-1} M V \equiv \left(\begin{array}{cc}
                               \lambda_{1}     & 0 \\
                                 0 & \lambda_{2}			 
									 \end{array}\right).
\end{eqnarray} 
Here the two eigenvalues $\lambda_{1}$ and $\lambda_{2}$ are specified as 
\begin{eqnarray}
 \lambda_{1}=\frac{a+b+D}{2} \ \ \mbox{and} \ \ \lambda_{2}=\frac{a+b-D}{2}.  \label{12i}
\end{eqnarray}

We can integrate Eq. (\ref{bold}) in the two cases, (i)$T > M$ and (ii)$ T < M$.
\begin{itemize}
\item[(i)]$T > M$  \par
The equation for the eigen-mode $X$ in Eq. (\ref{bold}) is integrated as
\begin{eqnarray}
 \frac{dX}{dT} &=& \lambda_{1} \cdot \gamma \frac{M^{3}}{T^{4}} X  \\
  \Leftrightarrow \ \ X(T)&=&X(T_{i}) \exp\biggl[-\frac{1}{3} \gamma \lambda_{1} 
  M^{3}\biggl(\frac{1}{T^{3}}-\frac{1}{T^{3}_{i}} \biggr)\biggr].   \label{sol1}
\end{eqnarray} 
It is natural for us to assume that $M \ll T_{i}$, where $T_{i}$ is the very high temperature
in the very early universe, for example, $T_{i} \simeq 10^{15}$GeV.  Then, the solution (\ref{sol1})
is written as
\begin{eqnarray}
 X(T=M)=X(T_{i}) e^{-\frac{1}{3} \lambda_{1} \gamma }.   \label{sol12}
\end{eqnarray} 

\item[(ii)]$T < M$  \par
The equation for $X$ is integrated as
\begin{eqnarray}
&& \frac{dX}{dT} = \lambda_{1} \cdot \delta \frac{M^{7/2}}{T^{9/2}} X  \\
  \Leftrightarrow && \ \log \frac{X(T_{f})}{X(T_{1})} = \lambda_{1} \delta M^{7/2}
 \int_{T_{1}}^{T_{f}} \frac{e^{M/T}}{T^{9/2}}dT  \\
    \Rightarrow && \  X(T_{f} \simeq 0)=X(T_{1}=M) e^{-3.19 \delta\lambda_{1}  },   \label{sol2}
\end{eqnarray} 
where  we set $T_{f}\simeq 0$, which is about the present temperature in the universe, 
and  $T_{1}=M$.
\end{itemize} 
We combine the solution (\ref{sol12}) with solution (\ref{sol2}) at the temperature $T=M$ and
obtain the time evolution of X from $T_{i}$ to $T_{f} \simeq 0$ as
\begin{eqnarray}
 X(T_{f} \simeq 0)=X(T_{i}) e^{-r\lambda_{1}},  \label{solt}
\end{eqnarray}
where 
\begin{eqnarray}
r\equiv \frac{1}{3}\gamma+3.19\delta \simeq 9.77.
\end{eqnarray}
The equation for another eigen-mode $Y$ in Eq. (\ref{bold}) is similarly integrated and 
Eq. (\ref{bold}) is solved in a simple form,
\begin{eqnarray}
\vec{N}_{f}^{'}=e^{-r\hat{M}}\vec{N}_{i}^{'}.  \label{sold}
\end{eqnarray}
When we go back to the original base $\vec{N}$, the solution is given as 
\begin{eqnarray}
\vec{N}_{f}=\left(\begin{array}{c}
                                   L_{f} \\
                                     \Delta_{f} 						 
									 \end{array}\right)
									 =V  \left(\begin{array}{cc}
                                  e^{-r\lambda_{1}} &  0 \\
                                  0  & e^{-r\lambda_{2}}
									 \end{array}\right) V^{-1}\left(\begin{array}{c}
                                   L_{i} \\
                                     \Delta_{i} 						 
									 \end{array}\right),
\end{eqnarray}
where the evolution matrix $S$ is concretely calculated to be 
\begin{eqnarray}
S  &\equiv& V e^{-r\hat{M}}  V^{-1} \\
&=& {\small \frac{1}{2D}\left(\begin{array}{cc}
                             (D+a-b)  e^{-r\lambda_{1}}+(D-a+b)  e^{-r\lambda_{2}}    & 
							 2d(e^{-r\lambda_{1}}-e^{-r\lambda_{2}})
							 \\
                              2c(e^{-r\lambda_{1}}-e^{-r\lambda_{2}})  &
							 (D-a+b)   e^{-r\lambda_{1}}+ (D+a-b)  e^{-r\lambda_{2}}  
									 \end{array}\right).} \nonumber
\end{eqnarray}
Thus the final lepton and triplet Higgs numbers are obtained as
\begin{eqnarray}
L_{f} =\frac{1}{2D} \biggl[ \bigl[ (D+a-b)L_{i}+2d \Delta_{i} \bigr]e^{-r\lambda_{1}}  
  + \bigl[(D-a+b)L_{i}-2d \Delta_{i} \bigr]e^{-r\lambda_{2}}   \biggr], \label{fln} \\
  \Delta_{f}=\frac{1}{2D}\biggl[ \bigl[ 2c L_{i}+(D-a+b) \Delta_{i} \bigr]e^{-r\lambda_{1}}  
  + \bigl[-2c L_{i}+(D+a-b)\Delta_{i} \bigr]e^{-r\lambda_{2}}   \biggr].  \label{fdn} 
 \end{eqnarray}
Using Eq. (\ref{bnum}) at the temperature $T_{f}$, we obtain the final baryon number,
\begin{eqnarray}				
B_{f} = \frac{1}{2D} \Biggl[ \biggl[\Bigl(-\frac{28}{51}(D+a-b)+\frac{4}{17}c \Bigr)L_{i}
+ \Bigl( -\frac{56}{51}d+\frac{2}{17}(D-a+b) \Bigr) \biggr]e^{-r\lambda_{1}}   
  \nonumber \\
  \hspace{2mm} +\Bigl[\biggl( -\frac{28}{51}(D-a+b)-\frac{4}{17}c \Bigr)L_{i}
+\Bigl( \frac{56}{51}d+\frac{2}{17}(D+a-b) \Bigr)\Delta_{i}  \biggr] e^{-r\lambda_{2}}	\Biggr].	
\label{fbn}
\end{eqnarray}

It should be noted that the determinant and trace of the matrix $M$ are both positive,
\begin{eqnarray}
\det(M)&=&\frac{289}{882} \cdot K_{f}\cdot K_{A}=\lambda_{1}\cdot \lambda_{2} > 0 \label{detm}  \\
  \mbox{and} \ \ Tr(M)&=&\frac{67}{28}K_{f}+\frac{5}{36}K_{A}=\lambda_{1}+\lambda_{2} >0,
  \label{trm}
\end{eqnarray}
which means that the both of the two eigenvalues, $\lambda_{1}$ and $\lambda_{2}$, are positive.
The positive eigenvalues tell us that the lepton number violating processes never generate the baryon number.
Eq. (\ref{detm}) also tells us that if either of $K_{f}$ or $K_{A}$ is zero, the determinant of $M$ is vanishing,
that is, $\lambda_{2} $=0. The vanishing of the second eigenvalue warrants the quantity, 
$P \equiv Y \varpropto ( bL-d \Delta )$, which is an eigen-mode in Eq. (\ref{eigenxy}),
to be conserved  and guarantees that the initial baryon number
surely survives in proportion to the conserved $U(1)$ charge $P$ as
\begin{eqnarray}
K_{f}=0 \ \ \mbox{or} \ \  K_{A}=0  \ \ &\Leftrightarrow&   \ \ \det(M)=0  \nonumber \\
&\Leftrightarrow& \ \  \lambda_{1}=a+b, \ \ \lambda_{2}=0  \nonumber \\
&\Rightarrow& \ \ B_{fin} \varpropto P=Y \varpropto ( bL-d \Delta ).      \label{exact}
\end{eqnarray}
In each of the two cases, (1)$K_{A}=0$ or (2)$ K_{f}=0$, the conserved $U(1)$ charge $P$ 
concretely becomes the following,
\begin{eqnarray}
\left\{\begin{array}{ll}
(1) \ K_{A}=0   & \Rightarrow \ \  P=L-2\Delta   \\
(2) \ K_{f}=0  & \Rightarrow \ \  P=L
\end{array}\right.
\end{eqnarray}

\subsection{Condition to protect the primordial baryon number  \label{con}}
\begin{figure}[ht]
 \begin{center}
 \includegraphics[width=8cm]{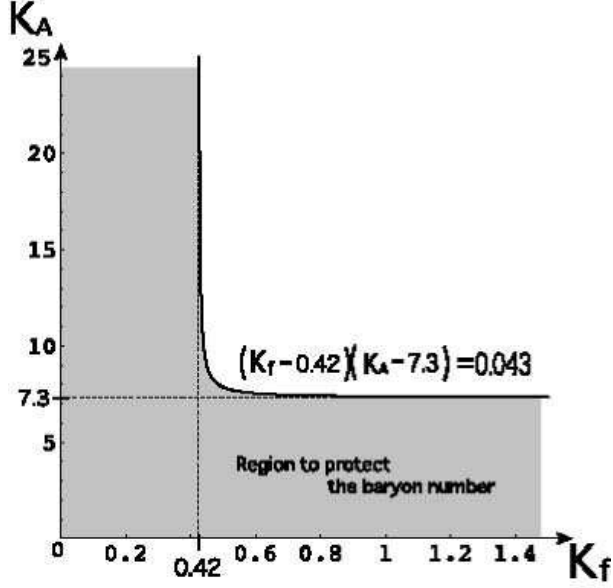}
 \end{center}
 \caption{
 The region where the baryon number is not washed out is shown as a gray region. 
Here  we require that the condition to protect the baryon number is $\lambda_{2}<1$.
\label{region1}  }
\end{figure}

In this subsection, we analyze the solutions (\ref{fln}), (\ref{fdn}), and (\ref{fbn})
 of the Boltzmann equation and obtain the condition that the initial baryon number is not washed out. 
 From the solutions  (\ref{fln}), (\ref{fdn}), and (\ref{fbn}) we immediately find that 
 the condition, $\lambda_{1}<1$ and $\lambda_{2}<1$ approximately conserves the lepton, triplet Higgs, and 
 baryon numbers as
 \begin{eqnarray}
\lambda_{1} < 1\  \mbox{and} \ \lambda_{2}<1\   \Rightarrow \ \ L_{f} \simeq L_{i},
\ \Delta_{f} \simeq \Delta_{i},  \ \mbox{and}   \ B_{f} \simeq B_{i}, 
\end{eqnarray}
where the two approximately conserved charges, $P=L$ and $\Delta$ independently exist.
 However in order to protect the baryon number,  two conserved charges
 are not needed. The only one conserved charge is enough to protect the 
 baryon number. 
 Hence the condition that the initial baryon number is not washed out is the following,
 \begin{eqnarray}
\lambda_{1} < 1\  \mbox{or} \ \lambda_{2}<1\   \Leftrightarrow \ \  \lambda_{2}<1 \ \ 
\Leftrightarrow \ \ B_{f}  \varpropto P=( bL-d \Delta )  \varpropto Y,
\end{eqnarray} 
 where we use the relation, $\lambda_{1} > \lambda_{2}$ in Eq. (\ref{12i}) and 
 the quantity $P$ is approximately conserved. Using $K_{f}$ and $K_{A}$, we 
 rewrite the condition, $\lambda_{2}<1$ as
  \begin{eqnarray}
 \lambda_{2}<1\   \Leftrightarrow \ \ \frac{a+b-D}{2} < 1 \ 
 \Leftrightarrow \ Tr(M)-2 < \sqrt{[ Tr(M)]^{2}-4\det(M)},  \label{bpc}
\end{eqnarray} 
 where $\det(M)$ and $Tr(M)$ are defined in Eq. (\ref{detm}) and (\ref{trm}). Inequality (\ref{bpc})
 is solved in the following two cases, (i)$Tr(M)-2<0$ or (ii)$Tr(M)-2>0$.
\begin{itemize}
\item[(i)] $Tr(M)-2<0$  \par
The condition, $Tr(M)-2<0$ is transformed to
\begin{eqnarray}
\frac{67}{28}K_{f}+\frac{5}{36}K_{A}-2<0.  \label{regi1}
\end{eqnarray} 
\item[(ii)] $Tr(M)-2>0$  \par
 Inequality (\ref{bpc}) is transformed to 
 \begin{eqnarray} 
\frac{67}{28}K_{f}+\frac{5}{36}K_{A}-2>0 \ \ \  \mbox{and} \ \ \ 
\biggl(K_{f}-\frac{245}{578} \biggr)\cdot \biggl(K_{A}-\frac{4221}{578} \biggr)<c_{1},  \label{regi2}
\end{eqnarray}
where the constant $c_{1}$ is
\begin{eqnarray}
c_{1}\equiv \biggl(\frac{882}{289} \biggr)^{2} \biggl(\frac{335}{1008}-\frac{289}{882} 
\biggr) \simeq 0.043.  
\end{eqnarray}
\end{itemize}
Either (\ref{regi1}) or  (\ref{regi2}) is the solution of inequality (\ref{bpc}) and is
shown in Fig. \ref{region1}. 

\begin{figure}[ht]
 \begin{center}
 \includegraphics[width=8cm]{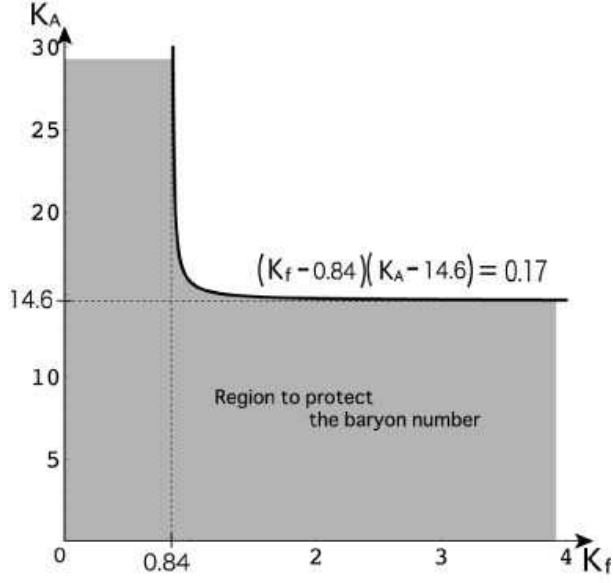}
 \end{center}
 \caption{
When we require that the condition to protect the baryon number is $\lambda_{2}<2$,
the allowed region is shown as a gray region.
\label{region2}  }
\end{figure}

 The condition $\lambda_{2}<1$ is rewritten for the damping factor as
\begin{eqnarray}
e^{-r\lambda_{2}}> e^{-9.77} \simeq 5.7 \times 10^{-5}.
\end{eqnarray} 
 If the initial baryon number is sufficiently large, the larger damping factor is allowed.
 For example, if we set the condition, $\lambda_{2}<2$ to protect the baryon number,
 the damping factor is the following,
 \begin{eqnarray}
e^{-r\lambda_{2}}> e^{-2 \times 9.77} \simeq 3.3 \times 10^{-9}.  \label{hard}
\end{eqnarray} 
If the initial baron number is $B_{i} \simeq 1$ as in the Affleck-Dine scenario,
the hard damping (\ref{hard}) 
can realize that the present baryon number, $B_{0} \simeq 10^{-10}$.
The condition $\lambda_{2}<2$ is 
rewritten as
   \begin{eqnarray}
 Tr(M)-4 < \sqrt{[ Tr(M)]^{2}-4\det(M)},  \label{bpc2}
\end{eqnarray} 
 and the allowed region for $K_{f}$ and $K_{A}$ is shown in Fig. \ref{region2}
 It is appropriate that we set the condition to protect the baryon number to be
    \begin{eqnarray}
K_{f} < 1 \ \ \mbox{or} \ \  K_{A} <1,  \label{bpcf}
\end{eqnarray}
 for the various damping factor.
 When this condition is satisfied, the quantity $P=bL-d\Delta$ is approximately conserved.
 The relation between the condition (\ref{bpcf}) and the conserved number $P=bL-d\Delta$ is consistent
 with the relation (\ref{exact}) in the limit $K_{f} \to 0$ or $K_{A} \to 0$. 
 In the limit $K_{f} \to 0$ or $K_{A} \to 0$, the approximately conserved number $P$
 becomes the exact conserved number associated with the exact symmetry $U(1)_{P}$.
 
 We next investigate how the final baryon number depends on $K_{f}$ and $K_{A}$
under the fixed initial conditions, $L_{i}$ and $\Delta_{i}$ and confirm the validity of the condition
(\ref{bpcf}).
 
\begin{figure}
 \begin{center}
 \includegraphics[width=11cm]{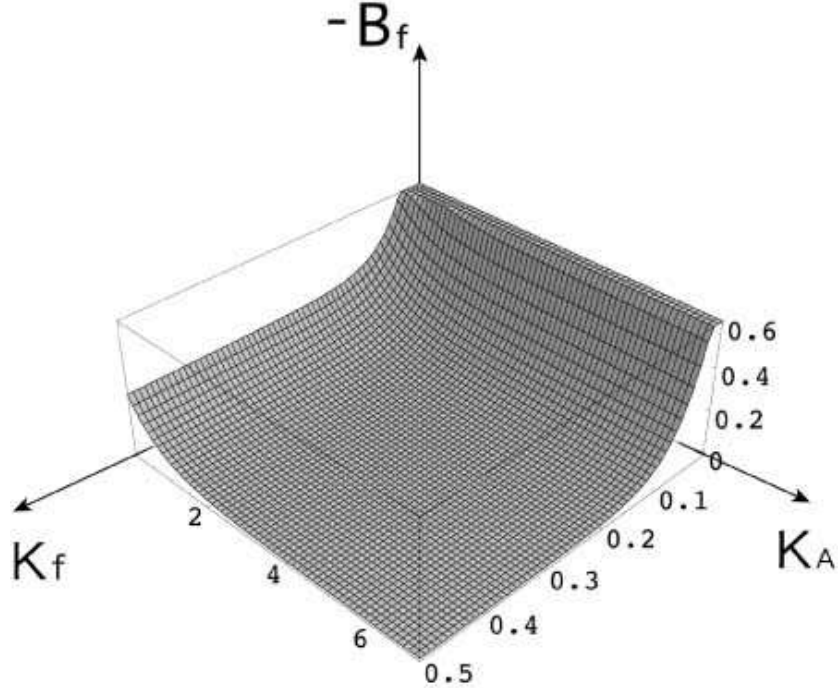}
 \end{center}
 \caption{The dependence of the final baryon number on $K_{f}$ and $K_{A}$
 is  shown when the initial condition $L_{i}=1$ and $\Delta_{i}=0$
is selected.
 \label{fb1} }
\end{figure}

\begin{figure}
 \begin{center}
 \includegraphics[width=11cm]{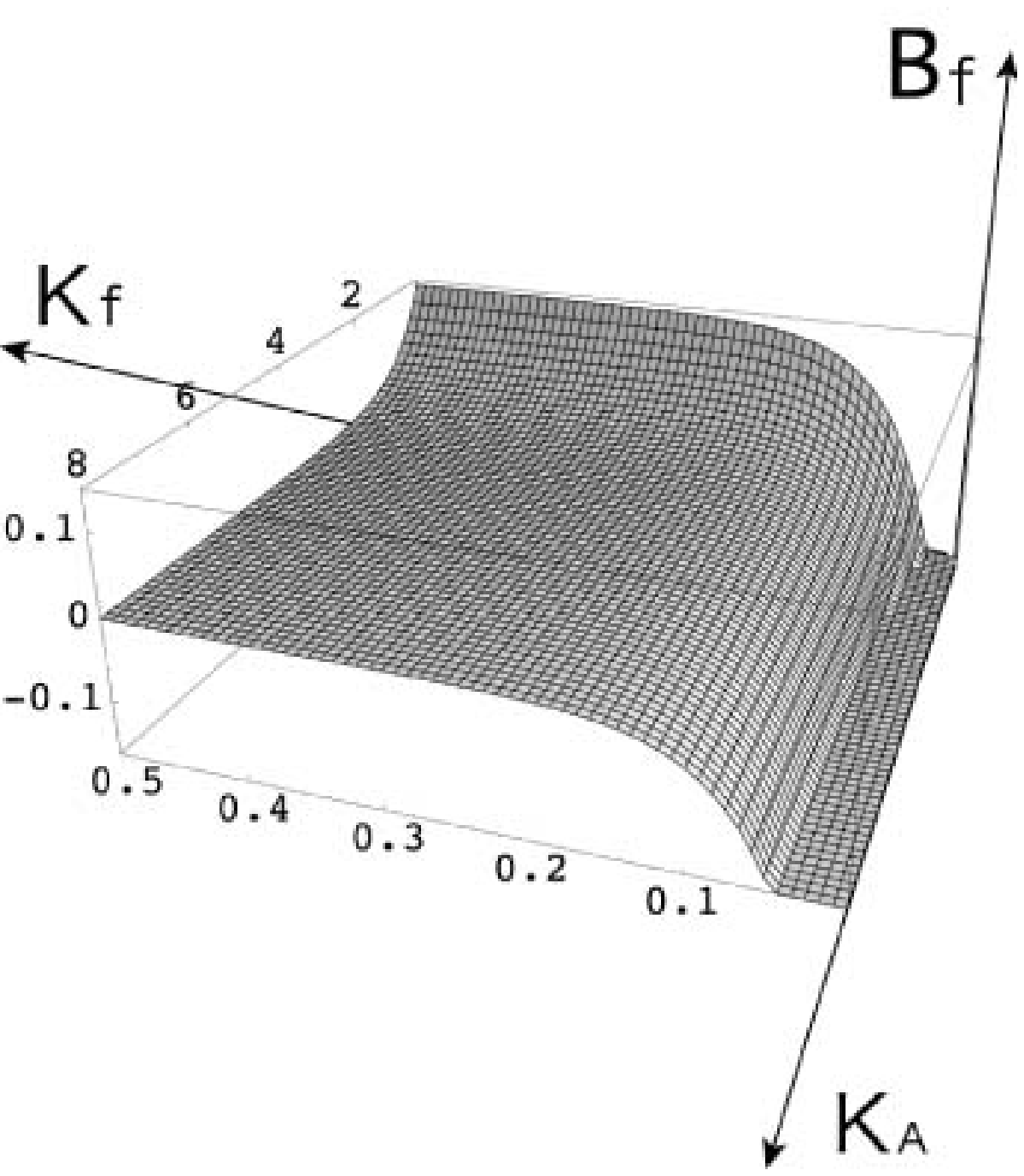}
 \end{center}
 \caption{The dependence of the final baryon number on $K_{f}$ and $K_{A}$
 is  shown when the initial condition $L_{i}=1$ and $\Delta_{i}=1$
is selected. \label{fb2} }
\end{figure}

 We first observe the case that the initial condition is 
     \begin{eqnarray}
 L_{i}=1  \ \ \mbox{and} \ \ \Delta_{i}=0.   \label{ini1}
\end{eqnarray}
Under the initial condition (\ref{ini1}), the final baryon number (\ref{fbn})
depends on $K_{f}$ and $K_{A}$ as in Fig. \ref{fb1}. 
It is found that the baryon number is not washed out in the region $K_{f}<1$ or $K_{A}<1$
in Fig. \ref{fb1}.
We second observe the case that the initial condition is 
     \begin{eqnarray}
 L_{i}=1  \ \ \mbox{and} \ \ \Delta_{i}=1.   \label{ini2}
\end{eqnarray}
Under the initial condition (\ref{ini2}), the final baryon number (\ref{fbn})
 depends on $K_{f}$ and $K_{A}$ as in Fig. \ref{fb2}.  
In this case, we again find that the baryon number is not washed out in the region $K_{f}<1$ or $K_{A}<1$
in Fig. \ref{fb2}. 

\begin{figure}
 \begin{center}
 \includegraphics[width=10cm]{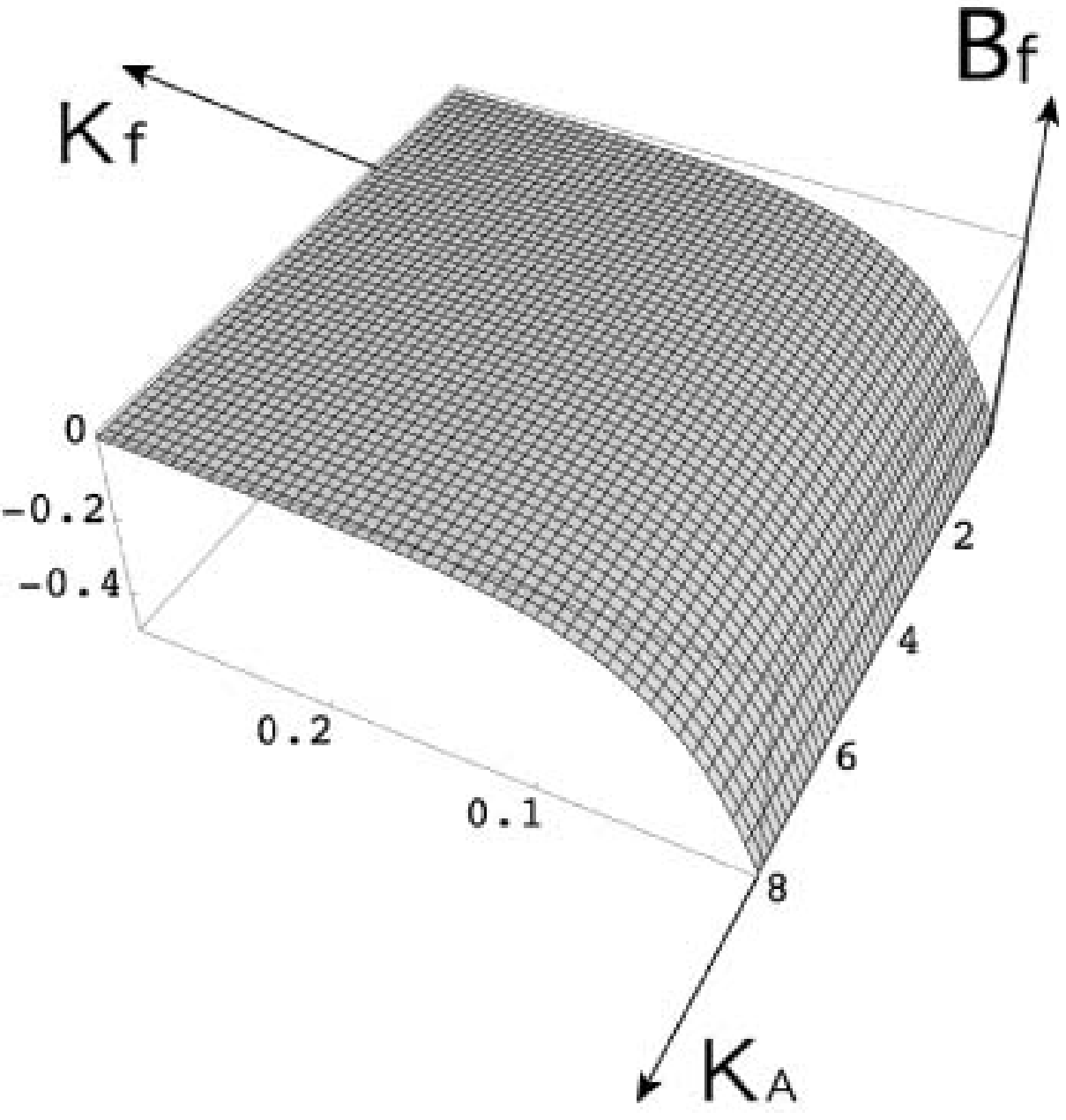}
 \end{center}
 \caption{The dependence of the final baryon number on $K_{f}$ and $K_{A}$
 is  shown when the initial condition $L_{i}=1$ and $\Delta_{i}=1/2$
is selected. \label{fb3} }
\end{figure}

We finally study the case that the initial condition is 
     \begin{eqnarray}
 L_{i}=1  \ \ \mbox{and} \ \ \Delta_{i}=\frac{1}{2}.  \label{ini3}
\end{eqnarray}
Under the initial condition (\ref{ini3}), the final baryon number (\ref{fbn})
 depends on $K_{f}$ and $K_{A}$ as in Fig. \ref{fb3}.  
In the figure, we find the anomalous behaver of the final baryon number.
It is shown that the baryon number is washed out at the region $K_{A}=0$, where the baryon
number should be protected. The reason is that the initial condition (\ref{ini3}) corresponds
to vanishing the quantity $P=L-2\Delta$ which becomes a conserved charge  at $K_{A}=0$ 
and the final baryon number 
is proportional to $P$, that is vanishing. The similar behaver occurs when the initial condition,
$L_{i}=0$ and $\Delta_{i}=1$ is selected. In this case,  the baryon number is washed out at
 the region $K_{f}=0$ because the initial condition erases the quantity,
 $P=L$ which becomes a conserved charge  at $K_{f}=0$. However, since the solutions 
(\ref{fln}), (\ref{fdn}), and (\ref{fbn}) are 
not the exact ones of the Boltzmann equations (\ref{Levo}), (\ref{Bevo}), (\ref{Devo}), (\ref{Pevo}), 
and (\ref{Wevo}), the conserved charge $P=bL-d\Delta$ in (\ref{exact})
 is actually not the case. The true conserved charges are the following,
 \begin{eqnarray}
K_{A}&=&0 \ \ \Rightarrow \ \ P=B-L+2 \Delta,   \nonumber \\
K_{f}&=&0 \ \ \Rightarrow \ \ P=B-L,   \label{true}
\end{eqnarray}
which are consistent with the conserved ones in the Boltzmann equation (\ref{bol3})
for the numbers, $B, L,$ and $\Delta $. 
 The anomalous behaver found above actually should occur when a initial condition erases the
 true conserved charge $P$ in (\ref{true}).

 We here generalize the condition (\ref{bpcf}) in order that
 we can apply the condition that the baryon number
 is not washed out to the various mass models of Majorana neutrino.
 The condition that the baryon number is not washed out in the mass models of Majorana 
 neutrino under the in-equilibrium sphaleron processes is generally that if  the effective lagrangian
 density is constructed only by the in-equilibrium interactions in the lagrangian density
 of a model, the effective lagrangian density has at least one conserved charge $P$ 
 associated with a group $U(1)_{P}$ symmetry which contains the lepton number. 
 In other words, the condition is that when the out-of-equilibrium interactions are
 eliminated in the lagrangian density of a model, the lepton number is not explicitly
 broken in the truncated lagrangian density.

In the above discussions, the lepton flavor has been omitted  because of Eq. (\ref{emt}) for simplicity.
When we take the lepton flavors, $e,\mu, \tau$ into account, the condition that the baryon number
 is not washed out is extended. Since the sphaleron-anomaly processes exactly conserve the three
 independent charges,
  \begin{eqnarray}
  \frac{B}{3}-L_{e}, \ \    \frac{B}{3}-L_{\mu}, \ \  \mbox{and}  \ \ \frac{B}{3}-L_{\tau},  \label{threech}
\end{eqnarray}
 the condition becomes that when the out-of-equilibrium interactions are eliminated in the
 lagrangian density of a model, at least one linear combination of the three charges (\ref{threech})
 is not  explicitly broken in the truncated  lagrangian density. Here the number of a lepton flavor,
$L_{\alpha} (\alpha=e, \mu, \tau)$ is defined as
 \begin{eqnarray}
L_{\alpha} \equiv \biggl[\frac{n_{\nu_{\alpha L}}-n_{\bar{\nu}_{\alpha L}}}{s}
+\frac{n_{e_{\alpha L}}-n_{\bar{e}_{\alpha L}}}{s}+\frac{n_{e_{\alpha R}}-n_{\bar{e}_{\alpha R}}}{s}\biggr].
\end{eqnarray}
Thus when we take the lepton flavors, $e,\mu, \tau$ into account
and require the cases that the only one  charge is approximately conserved,
there are ten possible approximately conserved $U(1)$ charges in the $SU(2)_{L}$ 
triplet Higgs model. We summarize all ten possible $U(1)$ charges and the 
conditions to enhance the $U(1)$ charges approximately in Table \ref{ap}.
 \begin{table}[ht] \begin{tabular}{|c|l|l|}\hline
 &  \ Approximately Conserved Charge & \ Condition \\ \hline 
(0)  & \ \ $P_{0}=B-L+2\Delta$  & \   $K_{A}<1$ \\ \hline 
(1)  & \ \  $P_{1}^{a}=\frac{B}{3}-L_{e}$  & \  $K_{L_{e}}<1$  \\ \hline 	
(2)  & \ \  $P_{1}^{a}=\frac{B}{3}-L_{\mu}$  &  \  $K_{L_{\mu}}<1$ \\ \hline 
(3)  & \ \  $P_{1}^{a}=\frac{B}{3}-L_{\tau}$  &  \  $K_{L_{\tau}}<1$  \\ \hline 
(4)  & \ \  $P_{2}^{a}=L_{e}-L_{\mu}$  & \   $K_{L_{e}-L_{\mu}}<1$  \\ \hline	
(5)  & \ \ $ P_{2}^{b}=L_{e}-L_{\tau}$  & \   $K_{L_{e}-L_{\tau}}<1$  \\ \hline	
(6)  & \ \  $P_{2}^{c}=L_{\mu}-L_{\tau}$  & \   $K_{L_{\mu}-L_{\tau}}<1$   \\ \hline		
(7)  & \ \ $P_{3}^{a}=\frac{B}{3}+L_{e \mu}$ & \   $K_{L_{e \mu}}<1$  \\ \hline 
(8) & \ \ $P_{3}^{b}=\frac{B}{3}+L_{e \tau}$ & \   $K_{L_{e \tau}}<1$   \\ \hline 
(9) & \ \ $P_{3}^{c}=\frac{B}{3}+L_{\mu \tau}$ & \   $K_{L_{\mu \tau}}<1$   \\ \hline  
\end{tabular}
\caption{In the $SU(2)_{L}$ triplet Higgs model, all ten possible $U(1)$ charges 
and the conditions to enhance the approximate $U(1)$ charges are shown.
The condition $K_{L_{e}}<1$ means that when we write all the processes
which violate the charge $L_{e}$ as $\Gamma_{L_{e}}$, $\Gamma_{L_{e}}<H |_{T=M}$.
The other conditions have similar meanings. \label{ap}}
\end{table}
In order to explain the condition $K_{L_{e}}<1$ in Table \ref{ap}, we define $\Gamma_{L_{e}}$ as
the reaction rates of the processes where the charge $L_{e}$ is violated. 
For example, $\Gamma_{L_{e}}$ are the reaction rates of the processes as
$\xi^{0} \longleftrightarrow \bar{\nu}_{e L} +\bar{\nu}_{\mu L}$ and
$\xi^{+} \longleftrightarrow \bar{\nu}_{e L} +e^{+}$, 
which are all caused by the coupling constants $f^{e \alpha} \ (\alpha=e, \mu, \tau)$.
The condition $K_{L_{e}}<1$ means that $\Gamma_{L_{e}}<H |_{T=M}$.
The other conditions have similar meanings.
Here the charges $L_{e \mu}, L_{e \tau},$ and $L_{\mu \tau}$ are defined as 
 \begin{eqnarray}
L_{e \mu}=L_{\tau}-L_{e}-L_{\mu},  \ L_{e \tau}=L_{\mu}-L_{e}-L_{\tau}, \ 
\mbox{and} \   L_{\mu \tau}=L_{e}-L_{\mu}-L_{\tau}.
\end{eqnarray}
The condition that the initial baryon number is not washed out is modified from
the condition (\ref{bpcf}) into 
\begin{eqnarray}
\mbox{"at least one of the ten conditions in Table \ref{ap} is satisfied in the model".}  \label{tencon}
\end{eqnarray}

\section{$SU(2)_{L}$ triplet HIggs model compatible with KamLAND and WMAP
\label{triprotect}}
We obtain the conditions (\ref{tencon}) that the primordial baryon number 
is not washed out in the $SU(2)_{L}$ triplet Higgs model. 
In this section, we investigate whether the constrained model by the conditions (\ref{tencon})
simultaneously satisfies the recent results of the neutrino oscillation experiments and WMAP, and
 the constraints on 
the $\rho$ parameter in the LEP.
In subsection \ref{status}, we first summarize the present status of the neutrino oscillation 
experiments.
In subsection \ref{ka}, we examine the condition (0) $K_{A}<1$ in ten conditions in 
Table \ref{ap}.
In subsection \ref{kf}, we thirdly examine the other nine conditions (1) $\sim$ (9) in Table \ref{ap}.
In subsection \ref{allow}, we finally  summarize the results of the present section and show the allowed
and forbidden regions of the parameters in the model.

\subsection{Present status of the results from the neutrino oscillation experiments \label{status}}
We summarize the present status of the neutrino oscillation experiments to make
clear what constraints the results of the experiments give to the neutrino mass matrix.
The Super-Kamiokande Collaboration(S-K) shows that there exist a mass squared difference
$\Delta_{a}$ and a mixing angle $\theta_{atm}$ in order to explain the
atmospheric neutrino oscillation \cite{S-K,S-K2,Valle-Gon},
  \begin{eqnarray}
 1.6 \times 10^{-3} \ \mbox{eV}^2 < \Delta_{a} < 4.0 \times 10^{-3} \ \mbox{eV}^2, \ \ 
0.88 < \sin^2{2\theta_{atm}} \le 1.0  \ (90 \% \ \mbox{C.L.}),   \label{SK}
\end{eqnarray}
with the best fit values $\Delta_{a}=2.5 \times 10^{-3} \ \mbox{eV}^2$
and $\sin^2{2\theta_{atm}}=1.00$. 
The global analysis of the first results of KamLAND combined with the existing data of
solar neutrino experiments shows that there exist a mass squared
difference $\Delta_{s}$ and a mixing angle $\theta_{\odot}$ to account for the solar neutrino oscillation
\cite{SNO,kam,SNO2,kam2,Valle-kam},
\begin{eqnarray}
5.1 \times 10^{-5} \ \mbox{eV}^2 < \Delta_{s} < 9.7 \times 10^{-5} \ \mbox{eV}^2&,& \  \ 
1.2 \times 10^{-4} \ \mbox{eV}^2 < \Delta_{s} < 1.9 \times 10^{-4} \ \mbox{eV}^2, \label{two} \nonumber
\\ 0.29 < \tan^2{\theta_{\odot}} <0.86 && (3 \sigma \ \mbox{level}), \label{non}
\end{eqnarray}
with the best fit values, $\Delta_{s}=6.9 \times 10^{-5}
\ \mbox{eV}^2$ and $\tan^2{\theta_{\odot}}=0.46$.
The CHOOZ experiment has put the upper bounds on the mixing angle $\theta_{13}$ \cite{Chooz},
\begin{eqnarray}
\sin^2{2\theta_{13}} \le 0.1. \label{CH} 
\end{eqnarray}

Using the Maki-Nakagawa-Sakata (MNS) matrix $U$,
we can diagonalize a Majorana neutrino mass matrix ${\cal M}_{\nu}$ to $\hat{\cal M}_{\nu}$,
\begin{eqnarray}
 \hat{\cal M}_{\nu} =\left(\begin{array}{ccc}
                                    m_{1} & 0 & 0 \\
                                     0  & m_{2} & 0 \\
									 0 & 0 & m_{3}
									 \end{array}\right)=  U^{T}{\cal M}_{\nu}U,  \hspace{15mm}
 \vec{\nu}_{L}^{m} =  U^{\dagger} \vec{\nu}_{L}\equiv \left(\begin{array}{c}
    	\nu_{1}\\   
       \nu_{2}  \\ 
         \nu_{3}
	\end{array}\right)_{L},
\end{eqnarray}
where the weak eigenstates in $\vec{\nu}_{L}$ are understood to be the partners of the mass eigenstates of
the charged leptons and $\vec{\nu}_{L}^{m}$ contain the neutrino mass eigenstates.
Since  we assume that ${\cal M}_{\nu}$ is a real symmetric matrix,
 we can parameterize the MNS matrix $U$ as a orthogonal matrix,
\begin{eqnarray}
 U &\equiv&  
   \left(\begin{array}{ccc}
    1   &   0   & 0    \\   
    0   &   c_{23}   &  s_{23}     \\ 
   0   &   -s_{23}    &  c_{23}     
	\end{array}\right)
 \left(\begin{array}{ccc}
    c_{13}   &   0   & s_{13}    \\   
    0   &   1   &  0      \\ 
   -s_{13}   &   0    &  c_{13}     
	\end{array}\right)
 \left(\begin{array}{ccc}
    c_{12}   &   s_{12}   & 0    \\   
    -s_{12}   &   c_{12}  &  0      \\ 
   0  &   0    &  1     
	\end{array}\right)  \nonumber \\
	&=& R_{23}(-\theta_{23})R_{13}(\theta_{13})R_{12}(-\theta_{12}), 
\end{eqnarray}
where $c_{ij} \equiv \cos \theta_{ij}$ and $s_{ij} \equiv \sin \theta_{ij}$. 
Using these parameters, the two mass squared differences and two mixing angles in Eq. (\ref{SK}) and (\ref{non})
are written as 
 \begin{eqnarray}
\Delta_{a}=|m_{3}^{2}-m_{2}^{2}|, \hspace{5mm} \Delta_{s}=m_{2}^{2}-m_{1}^{2}, \hspace{10mm}
\theta_{atm}=\theta_{23}, \hspace{5mm} \theta_{\odot}=\theta_{12}. \label{para}
 \end{eqnarray}
 The experimental data \cite{SNO,kam,SNO2,kam2,Valle-kam} for the solar neutrinos also show
 that the sign of $m_{2}^{2}-m_{1}^{2}$ is positive.
For brevity, we assume the mixing angle $\theta_{13}=0$, consistent with (\ref{CH}), 
and fix the two mass squared differences and ${\theta} _{atm}$ at the best fit values.
 Under these assumptions, we can write the matrix $U$ using the only one 
 relatively poorly known parameter $\theta_{\odot}\equiv \theta$,
  \begin{eqnarray}
 U =\left(\begin{array}{ccc}
    \cos \theta   &   \sin \theta   & 0    \\   
    -\frac{1}{\sqrt{2}}\sin \theta    &   \frac{1}{\sqrt{2}}\cos \theta   &  \frac{1}{\sqrt{2}}     \\
   \frac{1}{\sqrt{2}}\sin \theta   &   -\frac{1}{\sqrt{2}}\cos \theta   &     \frac{1}{\sqrt{2}}
	\end{array}\right)
	 \hspace{5mm} \mbox{with}  \hspace{3mm}   0.29 < \tan^2\theta <0.86  \ (3 \sigma \ \mbox{level}).
 \end{eqnarray}
 
Then, the neutrino mass matrix ${\cal M}_{\nu}$ can be written as  
 \begin{eqnarray}
 {\cal M}_{\nu}=\left(\begin{array}{ccc}
    c^{2}m_{1}+s^{2}m_{2}   &  -\frac{1}{\sqrt{2}}sc(m_{1}-m_{2})    &   \frac{1}{\sqrt{2}}sc(m_{1}-m_{2})   \\   
      -\frac{1}{\sqrt{2}}sc(m_{1}-m_{2})  &   \frac{1}{2}(s^{2}m_{1}+c^{2}m_{2}+m_{3})  &  
	  -\frac{1}{2}(s^{2}m_{1}+c^{2}m_{2}-m_{3})   \\ 
   \frac{1}{\sqrt{2}}sc(m_{1}-m_{2})   &   -\frac{1}{2}(s^{2}m_{1}+c^{2}m_{2}-m_{3}) & 
   \frac{1}{2}(s^{2}m_{1}+c^{2}m_{2}+m_{3})
	\end{array}\right), \label{m}
 \end{eqnarray}
where  $c \equiv \cos \theta$ and $s \equiv \sin \theta$. Since we have fixed
 the values of the two mass squared differences, there remains only one independent parameter
 among three mass eigenvalues, $m_{1}, m_{2}, m_{3}$.

While the neutrino oscillation experiments do not determine the magnitudes themselves
of the neutrino masses as in Eq. (\ref{para}),
the investigation into the cosmic microwave background radiation (CMBR) by WMAP
recently put the following upper bound on the sum of the neutrino masses 
\cite{Wmap1,Wmap2,Han,Elg,Wmap3}, 
  \begin{eqnarray}
\Omega_{\nu}h^{2}=\frac{\sum_{i}|m_{i}|}{93.5 \ \mbox{eV}} < 0.0076 \ 
\ \Leftrightarrow  \ \sum_{i}|m_{i}| < 0.71 \ \mbox{eV}. \label{WMAP}
\end{eqnarray}
In the following analyses, we require that the constraints (\ref{m}) and (\ref{WMAP})
are satisfied in the model.

\subsection{First condition $K_{A}<1$  \label{ka}}
In this subsection, we examine whether the condition (0) $K_{A}<1$ 
of ten conditions is compatible with the results (\ref{m}) of the neutrino oscillation experiments,
the results (\ref{WMAP}) of WMAP, and the constraints (\ref{lep}) on the 
$\rho$ parameter.
The condition (0) $K_{A}<1$ in Table \ref{ap} is rewritten from Eq. (\ref{orcon}) as 
\begin{eqnarray}
 |A| < 12 \times \biggl(\frac{M}{\mbox{GeV}}\biggr)^{\frac{3}{2}} \   \mbox{eV}.  \label{ao2}
\end{eqnarray}
The constraints (\ref{lep}) on the $\rho$ parameter is written as 
\begin{eqnarray}
|A|< 0.03 \times \frac{2\sqrt{2} M^{2}}{v}. \label{rho2}
\end{eqnarray}
We compare the right hand side of (\ref{ao2}) with one of (\ref{rho2}) as
\begin{eqnarray}
12 \times \biggl(\frac{M}{\mbox{GeV}}\biggr)^{\frac{3}{2}} 
 \mbox{eV} \ &<& \ 0.03 \times \frac{2\sqrt{2} M^{2}}{v} \nonumber \\
\Leftrightarrow \ M \ &>& \ 1.2 \ \mbox{eV},
\end{eqnarray}  
which shows that (\ref{ao2}) is more severe condition than (\ref{rho2}) because of
$M \gg 1.2$ eV.
We can easily show that the condition (\ref{ao2}) is compatible with the results
(\ref{m}) of the neutrino oscillation experiments.
When we fix the mass $M$ at 100 GeV, for example, 
the condition (\ref{ao2}) becomes
\begin{eqnarray}
 A \ < 12  \ \mbox{KeV}.   \label{ao100}  
\end{eqnarray}
When we further assume that the mass spectrum is normal type and $m_{1}=0,
 m_{2}>0,$ and $m_{3}>0$
and fix the solar mixing angle, $\theta$ in (\ref{non}) at the best fit value,
 the mass matrix (\ref{m}) becomes
 \begin{eqnarray}
 {\cal M}_{\nu} &\simeq& \left(\begin{array}{ccc}
 s^{2}\sqrt{\Delta_{s}}   &  \frac{1}{\sqrt{2}}sc\sqrt{\Delta_{s}}     &   -\frac{1}{\sqrt{2}}sc \sqrt{\Delta_{s}}   \\   
    \frac{1}{\sqrt{2}}sc\sqrt{\Delta_{s}}      &  \frac{1}{2}\sqrt{\Delta_{a}} & \frac{1}{2}\sqrt{\Delta_{a}}
 \\ 
 -\frac{1}{\sqrt{2}}sc \sqrt{\Delta_{s}}    &  \frac{1}{2}\sqrt{\Delta_{a}} & \frac{1}{2}\sqrt{\Delta_{a}}
	\end{array}\right)  \nonumber \\
	&\simeq&  \left(\begin{array}{ccc}
2.6 \times 10^{-3} &   2.7 \times 10^{-3}    & - 2.7 \times 10^{-3}    \\   
     2.7 \times 10^{-3}      & 2.5 \times 10^{-2} & 2.5 \times 10^{-2}
 \\ 
 - 2.7 \times 10^{-3}    &  2.5 \times 10^{-2}  & 2.5 \times 10^{-2}
	\end{array}\right) \ \mbox{eV}. \label{m100}
 \end{eqnarray}
When we fix $A$ at 1 eV, which is consistent with the condition (\ref{ao100}),
$f^{\alpha \beta}$ is calculated from the equation, $m^{\alpha \beta}\simeq$
1.5 eV $\times f^{\alpha \beta}$ in (\ref{numasstri}) and (\ref{m100})  as
 \begin{eqnarray}
f^{\alpha \beta} \simeq \frac{ {\cal M}_{\nu}}{1.5 \ \mbox{eV}} \simeq
\left(\begin{array}{ccc}
1.7 \times 10^{-3} &   1.8 \times 10^{-3}    & - 1.8 \times 10^{-3}    \\   
     1.8 \times 10^{-3}      & 1.7 \times 10^{-2} & 1.7 \times 10^{-2} \\ 
 - 1.8 \times 10^{-3}    &  1.7 \times 10^{-2}  & 1.7 \times 10^{-2}
\end{array}\right). \label{f1}
\end{eqnarray}
Inversely, when we take $f^{\alpha \beta}$ as (\ref{f1}),
both of the mass matrix (\ref{m100}) and the condition $K_{A}<1$ are simultaneously
satisfied. 
It is also easily found that the mass matrix (\ref{m100}) satisfies
the result (\ref{WMAP}) of the WMAP because of the equations,
\begin{eqnarray}
\sum_{i} |m_{i}|=\sum_{i} m_{i}=Tr M_{\nu} \simeq 5.3 \times 10^{-2} \ \mbox{eV} < 0.71 \ \mbox{eV}.
\end{eqnarray}
In the case $K_{A}<1$, the approximately conserved charge is  from (\ref{true}),
\begin{eqnarray}
P =B-L+2\Delta
\end{eqnarray}
and the final baryon and lepton numbers are calculated from a equation, 
\begin{eqnarray}
\mu_{\xi^{0}}+2\mu_{\nu_{\alpha}}=0,
\end{eqnarray}
(\ref{mux}), (\ref{munu}), and $(\ref{bnum})$
and are protected in proportion to the initial value $P_{ini}$ as
 \begin{eqnarray}
B_{fin}=-\frac{28}{33}L_{fin}=\frac{28}{229}P_{ini}.  
 \end{eqnarray}

We next investigate the region of the coupling constants $f^{\alpha \beta}$ where
the results of the neutrino oscillation experiments are not satisfied under 
the condition $K_{A}<1$.
We use the boundary values of the other nine conditions (1) $\sim$ (9) in Table \ref{ap}
as the standards which classify the regions of the coupling constants $f^{\alpha \beta}$. 
We show below that the constrained neutrino mass matrix (\ref{m})
can't satisfy any one of the nine conditions
in addition to the condition $K_{A}<1$.
The nine conditions (1) $\sim$ (9) are written as a form, 
\begin{eqnarray}  
 |f^{\alpha \beta}| \ < \ 4.3 \times 10^{-9} \times \biggl(\frac{M}{GeV} \biggr)^{\frac{1}{2}}.
 \label{fout}
 \end{eqnarray}
 The condition (\ref{ao2}) and (\ref{fout}) derive the constraint on the magnitudes of 
 the elements of the neutrino mass matrix in Eq. (\ref{numasstri}) as  
\begin{eqnarray}  
 |m^{\alpha \beta}|=|f^{\alpha \beta}| \frac{|A|v^{2}}{4M^{2}}<7.9 
 \times 10^{-4} \ \mbox{eV} \ (\equiv S).
 \label{mmag}
 \end{eqnarray}
 If the mass matrixes are restricted by the inequality (\ref{mmag}) for
 all $\alpha=e, \mu, \tau$, the matrixes are not able to induce 
 the atmospheric mass squared difference, $\sqrt{\Delta_{a}}=5.0\times 10^{-2}$ eV
 in (\ref{SK}). We next impose the condition (\ref{ao2}) and the one of the left nine
 conditions in Table \ref{ap}.
 We enumerate the type of the mass matrixes $M_{\nu}$ which are associated with 
 the exact nine symmetries which  the approximate nine symmetries
 go to in the limit, $K_{L} \to 0$ as 
\begin{eqnarray}  
 &P_{1}^{a}=L_{e}&  \hspace{7mm} P_{1}^{b}=L_{\mu}  \hspace{12mm} P_{1}^{c}=L_{\tau} \\
 M_{\nu}=  &\left(\begin{array}{ccc}
    0   &   0   & 0    \\   
    0   &   \times   &   \times     \\ 
   0   &    \times    &   \times     
	\end{array}\right),&
	   \left(\begin{array}{ccc}
     \times   &   0   &  \times    \\   
    0   &   0   &  0     \\ 
    \times  &   0    &   \times     
	\end{array}\right), \hspace{3mm}
	 \left(\begin{array}{ccc}
    \times  &   \times  & 0    \\   
     \times   &    \times   &  0     \\ 
   0   &   0    &  0     
	\end{array}\right),  \nonumber
 \end{eqnarray}
\begin{eqnarray}  
&P_{2}^{a}=L_{e}-L_{\mu}&  \hspace{4mm} P_{2}^{b}=L_{e}-L_{\tau}  \hspace{4mm} 
P_{2}^{c}=L_{\mu} -L_{\tau} \\
 M_{\nu}=   &\left(\begin{array}{ccc}
    0   &   \times   & 0    \\   
    \times   &   0   &   0    \\ 
   0   &   0   &   \times     
	\end{array}\right),&
	   \left(\begin{array}{ccc}
    0   &   0   &  \times    \\   
    0   &   \times   &  0     \\ 
    \times  &   0    &   0     
	\end{array}\right), \hspace{3mm}
	 \left(\begin{array}{ccc}
    \times  &  0  & 0    \\   
     0  &    0   &  \times     \\ 
   0   &   \times    &  0     
	\end{array}\right),  \nonumber
\end{eqnarray}
 \begin{eqnarray}  
&P_{3}^{a}=L_{e}+L_{\mu}-L_{\tau}&  \hspace{4mm}
 P_{3}^{b}=L_{e}-L_{\mu}+L_{\tau}  \hspace{4mm} 
P_{3}^{c}=-L_{e}+L_{\mu} +L_{\tau} \\
 M_{\nu}=   &\left(\begin{array}{ccc}
    0   & 0   &   \times    \\   
   0   &   0   &     \times    \\ 
     \times   &     \times   &   0     
	\end{array}\right),& \hspace{5mm}
	   \left(\begin{array}{ccc}
    0   &     \times   &  0    \\   
      \times  &   0   &    \times   \\ 
   0 &     \times    &   0     
	\end{array}\right),  \hspace{10mm}
	 \left(\begin{array}{ccc}
   0  &    \times  &   \times    \\   
       \times  &    0   & 0    \\ 
     \times  &   0    &  0     
	\end{array}\right).  \nonumber
 \end{eqnarray}
Thus we examine whether the mass matrixes associated with 
the nine approximate symmetries can satisfy the results of the neutrino oscillation experiments
or not. 
We first examine the case that the mass matrixes have the normal mass spectra.
\begin{itemize}
\item[(i)] Normal mass spectrum : $|m_{1}| < |m_{2}| < |m_{3}|$  \\
Choosing $m_{1}$ as a parameter, we can express $m_{2}$ and $m_{3}$ as a function of $m_{1}$, 
 \begin{eqnarray}
m_{3}&=&\pm \sqrt{\Delta_{a}+\Delta_{s}+m_{1}^{2}}, \\
m_{2}&=&\pm \sqrt{\Delta_{s}+m_{1}^{2}}, \\
m_{1}&>&0,
 \end{eqnarray}
 where $m_{1}$ can be taken to be non-negative with no loss of generality 
 thanks to the freedom of the re-phasing, 
 $\vec{\nu}_{L}^{m}\prime=\pm i \vec{\nu}_{L}^{m}$.
 Every element of the neutrino mass matrix can be written as 
 \begin{eqnarray}
m^{ee}&=&c^{2}m_{1}\pm s^{2} \sqrt{\Delta_{s}+m_{1}^{2}}, \label{ee}  \\
m^{e\mu}&=&-m^{e\tau}=-\frac{1}{\sqrt{2}}sc\biggl(m_{1}\mp \sqrt{\Delta_{s}+m_{1}^{2}}\biggr) ,  \label{emu} \\
m^{\mu \tau}&=&-\frac{1}{2}\biggl(s^{2}m_{1}\pm c^{2}\sqrt{\Delta_{s}+m_{1}^{2}}\mp
\sqrt{\Delta_{a}+\Delta_{s}+m_{1}^{2}}\biggr), \label{mutau} \\
m^{\mu \mu}&=&m^{\tau \tau}=\frac{1}{2}\biggl(s^{2}m_{1}\pm c^{2}\sqrt{\Delta_{s}+m_{1}^{2}}\pm
\sqrt{\Delta_{a}+\Delta_{s}+m_{1}^{2}}\biggr). \label{mumu}
 \end{eqnarray}
For the approximate symmetries, $(P_{1}^{a},P_{2}^{a},P_{2}^{b},P_{3}^{a},P_{3}^{b},
P_{3}^{c})$,
$|m^{ee}|$ in Eq. (\ref{ee}) is restricted as
 \begin{eqnarray}  
|m^{ee}|=\biggl|c^{2}m_{1}\pm s^{2} \sqrt{\Delta_{s}+m_{1}^{2}} \biggr| < S=7.9 \times 10^{-4}
 \ \mbox{eV}.    \label{nineq1}
 \end{eqnarray}
 If $m_{2}>0$, Eq. (\ref{ee}) induces the inequality,
 \begin{eqnarray}
 m^{ee} \ge s^{2}\sqrt{\Delta_{s}} \ge s^{2}_{min} \sqrt{\Delta_{s}}=1.8 \times 10^{-3}
 \ \mbox{eV} >S. \label{nineq2}
 \end{eqnarray}
Inequalities (\ref{nineq1}) and (\ref{nineq2}) need the inequality $m_{2}<0$.
Here we shorten $\sin \theta_{min}$ as $s_{min}$  and 
$\theta_{min}$ is defined as $\tan^{2} \theta_{min}=0.29$ in Eq. (\ref{non}).
For $(P_{1}^{a},P_{2}^{a},P_{2}^{b},P_{3}^{a},P_{3}^{b})$, 
the element $ |m^{e\mu}|$ or $|m^{e\tau}|$
has to be restricted as
\begin{eqnarray}
|m^{e\mu}| < S \ \mbox{or} \ |m^{e\tau}| < S. \label{emus}
 \end{eqnarray}
On the other hand, Eq. (\ref{emu}) with $m_{2}<0$ derives the inequality, 
\begin{eqnarray}
 |m^{e\mu}|= |m^{e\tau}| &=&\frac{1}{\sqrt{2}}sc\biggl(m_{1}+\sqrt{\Delta_{s}+m_{1}^{2}}\biggr)
 \nonumber\\
 &\ge& \frac{1}{\sqrt{2}}s_{min}c_{min}\sqrt{\Delta_{s}}=2.4 \times 10^{-3} \ \mbox{eV} >S, \label{S1}
 \end{eqnarray}
 where $c_{min}\equiv \cos \theta_{min}$.
 Since (\ref{S1}) is inconsistent with (\ref{emus}), the approximate symmetries, 
 $(P_{1}^{a},P_{2}^{a},P_{2}^{b},P_{3}^{a},P_{3}^{b})$ are not allowed.
 For $P_{3}^{c}$, the inequality $|m^{\mu\mu}|<S$ is needed.
 On the other hand, Eq. (\ref{mumu}) derives the inequalities,
\begin{eqnarray}
 m_{3}>0 &\Leftrightarrow&  \ m^{\mu\mu} \ge 2.2 \times 10^{-2} \ \mbox{eV}  > S,  \\
 m_{3}<0 &\Leftrightarrow&  \ m^{\mu\mu} \le -2.8 \times 10^{-2} \ \mbox{eV} < -S.  
 \end{eqnarray}
 which don't allow the inequality $|m^{\mu\mu}|<S$, that is, the approximate symmetry $P_{5}^{c}$.
 The left  approximate symmetries, $(P_{1}^{b},P_{1}^{c},P_{2}^{c})$ need the inequality,
   \begin{eqnarray}
   |m^{e\mu}| < S \ \mbox{or} \ |m^{e\tau}| < S,
 \end{eqnarray}
which require $m_{2}>0$ because of (\ref{S1}). The symmetries further need the inequality,
\begin{eqnarray}
 |m^{\mu\mu}| < S \ \mbox{or} \ |m^{\tau \tau}| < S. \label{s1}
 \end{eqnarray}
On the other hand, Eq. (\ref{mumu}) with $m_{2}>0$ derives the inequality,
\begin{eqnarray}
 m_{3}>0 &\Rightarrow&  \ |m^{\mu\mu}|= |m^{\tau\tau}| \ge \frac{1}{2}\sqrt{\Delta_{a}}>S.  \label{s2}
 \end{eqnarray}
 Since the inequalities (\ref{s1}) and (\ref{s2}) are inconsistent, the inequality $m_{3}<0$ is needed.
 For $P_{1}^{b}$ and $P_{1}^{c}$, the inequality, $|m^{\mu\tau}|<S$ is needed.
 On the other hand, Eq. (\ref{mutau}) under the inequalities $m_{2}>0$ and $ m_{3}<0$ induce
 the inequality,
\begin{eqnarray}
|m^{\mu\tau}| \ge \frac{1}{2}\sqrt{\Delta_{a}}>S,
 \end{eqnarray}
 which don't allow $|m^{\mu\tau}|<S$, that is,  the approximate symmetries $P_{1}^{b}$ and $P_{1}^{c}$.
The left one approximate symmetry $P_{3}^{c}$ needs the inequality, $|m^{\mu\mu}|<S$
which is solved under $m_{2}>0$ and $m_{3}<0$ for $|m_{1}|$ in Eq. (\ref{mumu}) as
\begin{eqnarray}
 |m^{\mu\mu}|<S \Rightarrow  \ |m_{1}| \ge 0.80 \ \mbox{eV},  
 \end{eqnarray}
 which is inconsistent with the results (\ref{WMAP}) of WMAP  and don't allow the 
 approximate symmetry $P_{2}^{c}$. From the inconsistency obtained above, 
 we find that none of the nine approximate symmetries
 can be compatible with the neutrino mass matrixes with
 normal mass spectra under the condition $K_{A}<1$.

\item[(ii)] Inverted mass spectrum : $|m_{3}| < |m_{1}| < |m_{2}|$  \\
Choosing the smallest mass $m_{3}$ as a parameter, we can express $m_{1}$ and $m_{2}$ 
as a function of $m_{3}$, 
 \begin{eqnarray}
m_{2}&=&\pm \sqrt{\Delta_{a}+\Delta_{s}+m_{3}}, \\
m_{1}&=&\pm \sqrt{\Delta_{a}+m_{3}}, \\
m_{3} &>&0,
 \end{eqnarray}
 where $m_{3}$ can be taken to be non-negative without loss of generality.
  Every element of the neutrino mass matrix can be written as 
\begin{eqnarray}
m^{ee}&=&\pm c^{2}\sqrt{\Delta_{a}+m_{3}^{2}}\pm s^{2}\sqrt{\Delta_{a}+\Delta_{s}+m_{3}^{2}}, \label{iee} \\
m^{e\mu}&=&-m^{e\tau}=-\frac{1}{\sqrt{2}}sc\biggl(\pm \sqrt{\Delta_{a}+m_{3}^{2}}
\mp  \sqrt{\Delta_{a}+\Delta_{s}+m_{3}^{2}} \biggr), \label{iemu}\\
m^{\mu \tau}&=&-\frac{1}{2}\biggl(\pm s^{2}\sqrt{\Delta_{a}+m_{3}^{2}}\pm
 c^{2}\sqrt{\Delta_{a}+\Delta_{s}+m_{3}^{2}}-m_{3}\biggr), \label{imutau}\\
m^{\mu \mu}&=&m^{\tau \tau}=\frac{1}{2}\biggl(\pm s^{2}\sqrt{\Delta_{a}+m_{3}^{2}}
 \pm c^{2}\sqrt{\Delta_{a}+\Delta_{s}+m_{3}^{2}}+m_{3}\biggr). \label{imumu}
 \end{eqnarray} 
For the approximate symmetries
$(P_{1}^{a},P_{2}^{a},P_{2}^{b},P_{3}^{a},P_{3}^{b},P_{3}^{c})$,
the inequality $|m^{ee}|<S$ is needed. On the other hand,  $|m^{ee}|$ 
under the inequality, $m_{1} \cdot m_{2}>0$ in Eq. (\ref{iee}) derives the inequality,
\begin{eqnarray}
 |m^{ee}|\ge \sqrt{\Delta_{a}}>S,
 \end{eqnarray}
 which is inconsistent with $|m^{ee}|<S$ and require the inequality, $m_{1} \cdot m_{2}<0$.
In the  approximate symmetries $(P_{1}^{a},P_{2}^{a},P_{2}^{b},P_{3}^{a},P_{3}^{b})$,
the element $|m^{e\mu}|$ or $|m^{e\tau}|$ has to satisfy the inequality,  
  \begin{eqnarray}
 |m^{e\mu}| < S \ \mbox{or} \ |m^{e\tau}| < S. \label{emus2}
 \end{eqnarray}
 On the other hand, Eq. (\ref{iemu}) with the inequality, $m_{1} \cdot m_{2}<0$
 derive the inequality,
\begin{eqnarray}
 |m^{e\mu}|= |m^{e\tau}| \ge \sqrt{2} s_{min}c_{min}\sqrt{\Delta_{a}}
 =2.9 \times 10^{-2} \ \mbox{eV} >S,  \label{107}
 \end{eqnarray}
 which is inconsistent with the inequality (\ref{emus2}) and doesn't allow the 
 approximate symmetries $(P_{1}^{a},P_{2}^{a},P_{2}^{b},P_{3}^{a},P_{3}^{b})$.
 For $P_{3}^{c}$, the inequalities, $|m^{\mu\mu}|<S$ and $|m^{\mu\tau}|<S$ are needed.
 On the other hand, Eqs. (\ref{imutau}) and (\ref{imumu}) derive the inequalities,
\begin{eqnarray}
 m_{1}>0 \ \mbox{and} \ m_{2}<0 &\Rightarrow&  \ |m^{\mu\tau}| \ge 2.0 \times 10^{-2} \ \mbox{eV} > S,  \\
 m_{1}<0 \ \mbox{and} \ m_{2}>0 &\Rightarrow&  \ |m^{\mu\mu}| \ge 2.0  \times 10^{-2} \ \mbox{eV} >S,  
 \end{eqnarray}
 which are inconsistent with $|m^{\mu\mu}|<S$ and $|m^{\mu\tau}|<S$ and don't allow
the approximate symmetries $P_{3}^{c}$. The left symmetries $(P_{1}^{b},P_{1}^{c},P_{2}^{c})$ need
the inequalities,
\begin{eqnarray}
|m^{e\mu}| < S \ \mbox{or} \  |m^{e\tau}| < S,
\end{eqnarray}
which require the inequality, $m_{1} \cdot m_{2}>0 $ from the inequality (\ref{107}).
The symmetries $P_{1}^{b}$ and $P_{1}^{c}$ also need the inequalities,
\begin{eqnarray}
"|m^{\mu \tau}| < S  \ \ &\mbox{and}& \ \  |m^{\mu \mu}| \ < \ S" \nonumber \\
&\mbox{or}&  \nonumber \\
"|m^{\mu \tau}| < S  \ \ &\mbox{and}&  \ \ |m^{\tau \tau}| \ < \ S". \label{either}
\end{eqnarray}
On the other hand, Eq. (\ref{imumu}) and (\ref{imutau}) derive the inequalities,
\begin{eqnarray}
 m_{1}>0 \ \mbox{and} \ m_{2}>0 &\Rightarrow&  \ |m^{\mu\mu}|=|m^{\tau\tau}|
 \ge \frac{1}{2}\sqrt{\Delta_{a}} > S,  
 \label{mms} \\
 m_{1}<0 \ \mbox{and} \ m_{2}<0 &\Rightarrow&  \ |m^{\mu\tau}| \ge \frac{1}{2}\sqrt{\Delta_{a}} > S,  
 \end{eqnarray}
 which are inconsistent with the inequalities (\ref{either})
 and don't allow  the approximate symmetries $P_{1}^{b}$ and $P_{1}^{c}$.
 The left one approximate symmetry $P_{2}^{c}$ needs the inequalities, 
 $|m^{\mu\mu}|<S$ and $|m^{\tau\tau}|<S$, which derives the inequality,
 $m_{1}<0$ and $m_{2}<0$
 from (\ref{mms}). The inequality,  $|m^{\mu\mu}|<S$ with $m_{1}<0$ and $m_{2}<0$
 is solved for $m_{3}$ from (\ref{imumu}) as
 \begin{eqnarray}
 |m^{\mu\mu}|<S \Rightarrow  \ |m_{3}| \ge 0.81  \ \mbox{eV},  
 \end{eqnarray}
 which is inconsistent with the results (\ref{WMAP}) and don't allow the 
 approximate symmetry $P_{2}^{c}$.
 From the inconsistency above, we find that none of the nine approximate symmetries
 can be compatible with the mass matrixes with
 inverted mass spectra under the condition $K_{A}<1$.

\end{itemize}

The above analyses show that both of the condition $K_{A}<1$ and any one of the nine
conditions (1) $\sim$ (9) in Table \ref{ap} are not simultaneously
satisfied in the mass matrixes compatible with the results of the 
neutrino oscillation experiments.
Such results favor the relation
that the coupling constant $A$ is much smaller than the electroweak 
scale but the coupling constants $f^{\alpha \beta}$ are
the same order as the Yukawa couplings of the charged leptons in (\ref{cupmag}).
Such allowed region is expected to be natural in the $SU(2)_{L}$ triplet Higgs model
 in the framework of the large extra dimension.

\subsection{Other nine conditions \label{kf}}
In this subsection, we impose the other nine conditions (1) $\sim$ (9) in Table \ref{ap}
as
\begin{eqnarray}
|f^{\alpha \beta}| < 4.3 \times 10^{-9} \times   
 \biggl(\frac{M}{\mbox{GeV}}\biggr)^{\frac{1}{2}}. \label{foe2} 
\end{eqnarray}
Here, what $\alpha$ and $\beta$ satisfy the inequality (\ref{foe2})
depends on the condition which we select in the nine conditions.
Since in the previous subsection we find that both of the condition $K_{A}<1$ and any one of the nine
conditions are not simultaneously
satisfied in the mass matrixes compatible with the neutrino oscillation experiments,
under the condition (\ref{foe2}) we need to search for the parameter region $K_{A}>1$, that is,
  \begin{eqnarray}
  |A| > 12 \times \biggl(\frac{M}{\mbox{GeV}}\biggr)^{\frac{3}{2}} 
 \mbox{eV}.  \label{ae2}
\end{eqnarray}
The constraints (\ref{rho2}) on the $\rho$ parameter and the region (\ref{ae2})
are written as
 \begin{eqnarray}
12  \times \biggr( \frac{M}{GeV}\biggl)^{\frac{3}{2}}\ \mbox{eV} \ 
< \ |A| \ < 3.5 \times 10^{5} \times \biggr( \frac{M}{GeV}\biggl)^{2} \mbox{eV}. \label{abet}
 \end{eqnarray}
The inequalities (\ref{foe2}) and (\ref{rho2}) give the upper limit on the neutrino masses as 
\begin{eqnarray}
 |m^{\alpha \beta}|=|f^{\alpha \beta}| \frac{|A|v^{2}}{4M^{2}}< 26 \times 
 \biggr( \frac{M}{GeV}\biggl)^{\frac{1}{2}}
 \mbox{eV},  \label{mlim}
 \end{eqnarray}
 which seems to be compatible with the neutrino oscillation experiments.
 We actually show that both of the conditions, (\ref{foe2}) and (\ref{abet})
 are compatible with the results of the neutrino oscillation experiments
 in a typical example.
When we assume that the neutrino mass matrix is (\ref{m100}) 
and fix the mass $M$ at 100 GeV again, 
the conditions (\ref{foe2}) and (\ref{abet}) are written 
as 
\begin{eqnarray}
&& |f^{\alpha \beta}| \ < \ 4.3 \times 10^{-8} \label{foe100}, \\ 
&& 8.6  \ \mbox{KeV} \ < \ |A| \ < 3.5 \ \mbox{GeV}, \label{abet100}  
  \end{eqnarray}
  When we further fix the coupling constant $A$ at 100 MeV, the coupling constants $f^{\alpha \beta}$
are obtained from the equation, $m^{\alpha \beta} \simeq$ 151 MeV $\times 
f^{\alpha \beta}$ and (\ref{m100}) as
 \begin{eqnarray}
f^{\alpha \beta} \simeq \frac{ {\cal M}_{\nu}}{151 \ \mbox{MeV}} \simeq
\left(\begin{array}{ccc}
1.7 \times 10^{-11} &  1.8 \times 10^{-11}    & -1.8 \times 10^{-11}    \\   
     1.8 \times 10^{-11}      & 1.7 \times 10^{-10} & 1.7 \times 10^{-10}
 \\ 
 - 1.8 \times 10^{-11}    &  1.7 \times 10^{-10}  & 1.7 \times 10^{-10}
	\end{array}\right), \label{m1im}
 \end{eqnarray}
which are consistent with (\ref{foe100}) for all $\alpha$ and $\beta$.
Hence all the nine conditions (1) $\sim$ (9) are simultaneously
compatible with  the results of the neutrino oscillation experiments.
In the case that all the nine conditions are simultaneously satisfied,
the allowed approximately conserved charge is from (\ref{true}), 
\begin{eqnarray}
P = B-L
\end{eqnarray}
and the final baryon and lepton numbers are calculated from a equation, 
\begin{eqnarray}
2\mu_{\phi^{0}}-2\mu_{\xi^{0}}=0,
\end{eqnarray}
(\ref{mup}), (\ref{mux}), and $(\ref{bnum})$ and are protected in 
proportion to the initial value $P_{ini}$ as
 \begin{eqnarray}
B_{fin}=-\frac{52}{105}L_{fin}=\frac{52}{157}P_{ini}.  
 \end{eqnarray}

Since we expect that the small coupling $A$ and the Yukawa couplings $f^{\alpha \beta}$, being 
same order as those of the charged leptons, to be natural in this model, the conditions (\ref{foe100})
and (\ref{abet100}) are necessarily regarded as the unnatural parameter region in the model.

\subsection{Allowed region compatible with all requirements  \label{allow}}
\begin{figure}[ht]
 \begin{center}
 \includegraphics[width=12cm]{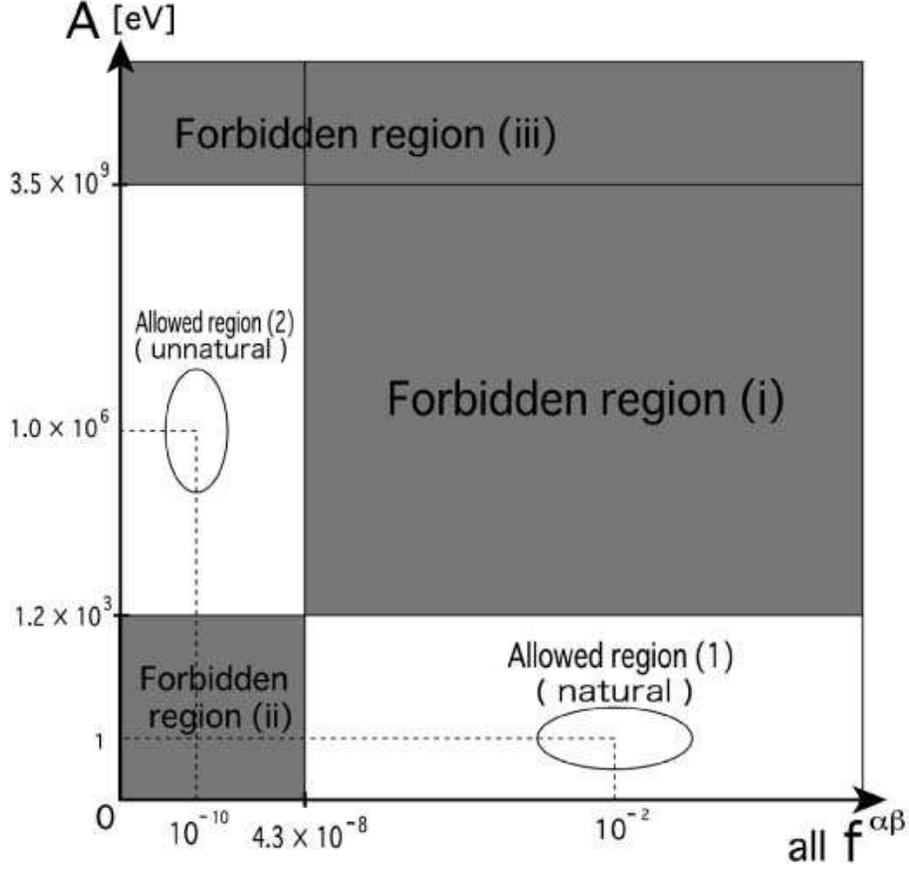}
 \end{center}
 \caption{Three forbidden regions (i), (ii),  and (iii) are shown as gray regions.
 Two allowed regions (1) and (2) which we confirm are shown as white regions.
 The masses of the triplet Higgs fields are fixed at $M=$100GeV.
 \label{asfig} }
\end{figure}

In Subsection \ref{ka} and \ref{kf}, we examine whether each of ten conditions in Table \ref{ap}
is compatible with the results of the neutrino oscillation experiments and WMAP, the constraints on the 
$\rho$ parameter and we obtain the allowed region of the parameter. 
Since the allowed region complicatedly depends on the six coupling constants $f^{\alpha \beta}$,
the  allowed region are not completely obtained. 
Here, for simplicity, we consider the only two cases in which all the coupling constants
$f^{\alpha \beta}$ satisfy the condition (\ref{foe2}) and in which none of $f^{\alpha \beta}$
satisfy  (\ref{foe2}). We further fix the masses of the triplet Higgs fields at $M=$100GeV.
Under the assumptions we show a typical allowed region in Fig. \ref{asfig}.
The three forbidden regions where the requirements are not satisfied are shown
 in the gray region in Fig. \ref{asfig}.
The conditions that the primordial baryon number is not washed out 
are not satisfied in the forbidden region (i) where
both of the coupling constants $A$ and $f^{\alpha \beta}$ are large.
The results of the neutrino oscillation experiments are not satisfied
 in the forbidden region (ii) where 
 both of the coupling constants $A$ and $f^{\alpha \beta}$ are small.
The constraint on the $\rho$ parameter is not satisfied
in the forbidden region (iii) where $A$ is large.
Although we exactly obtain the forbidden region,
we do not  exactly obtain the allowed region.
What we show is that there exist the allowed region in the two white regions in Fig. \ref{asfig}.
 The allowed region (1) in the white region where the coupling constant $A$ is the very small  and 
 the Yukawa couplings $f^{\alpha \beta}$ are as large as those of the charged leptons in (\ref{cupmag})
is natural parameter region of the $SU(2)_{L}$ triplet Higgs model in the large extra dimension.
 Inversely the allowed region (2) in the white region where $A$ is the large and
 $f^{\alpha \beta}$ are very small is expected to be unnatural.

\section{Summary    \label{trisum}}
\subsection{Conclusions}
We introduce the $SU(2)_{L}$ triplet Higgs model where the smallness of
the neutrino masses is naturally explained by the shining mechanism in the 
large extra dimension. In order to violate the lepton number explicitly,
which is the source of the Majorana mass terms,
we need to introduce the two kind of the interactions
in addition to the ones in the standard model.
It is a first fruitful consequence that we can obtain the condition (\ref{tencon})
that the primordial baryon number is not washed out by solving the Boltzmann
equations in the model.
We find that in order to protect the primordial baryon number,
either of the two kind of the interactions, both of which are needed to violate the
lepton number explicitly, must be out of equilibrium under the in-equilibrium 
sphaleron-anomaly processes.
The condition to protect the primordial baryon number is easily extended
into the case that the differences of the three lepton flavors are taken into account.
It is a second fruitful consequence that we can obtain the allowed region 
where the condition to protect the primordial baryon number, 
the results of the neutrino oscillation experiments and WMAP,
and the constraints on the $\rho$ parameter are all satisfied in the model.
A typical allowed region is shown in Fig. \ref{asfig}.
We find that if both of the coupling constant $A$ and $f^{\alpha \beta}$ are large,
the existing baryon number is washed out and when both of $A$ and $f^{\alpha \beta}$ 
are small, the mass differences which are induced from the neutrino mass matrix
are too small  to explain the observed neutrino oscillations.
In this model, the constraints on the $\rho$ parameter give the upper bounds 
on the coupling constant $A$ for the fixed triplet Higgs masses.
We find the two allowed regions where all the requirements are satisfied.  
One region where $A$ is very small and $f^{\alpha \beta}$ are as large as the 
Yukawa coupling constants of the charged leptons is expected
to be a favorable region where the smallness of the neutrino masses is naturally 
explained in the $SU(2)_{L}$ triplet Higgs model in the large extra dimension.

\subsection{Discussions}
We first make comment on the pre-existing approximately conserved 
charge. We actually obtain the approximately conserved charges,
$P=B-L+2\Delta$ or $P=B-L$. The primordial baryon number is protected in proportion
to the initial value of the conserved charge.
Can we naturally expect that the initial amount of the conserved charge exists
in the early universe ?
We recall that in the $SU(5)$ GUT, the initial value of the conserved charge
$(B-L)$ is naturally assumed to be zero, which regrettably leads to the washing out of the baryon
number by the in-equilibrium sphaleron processes. 
The difference between the case we examine and the case of $SU(5)$ GUT is that
in the former case the conserved charge is approximate one, while in the latter case
the conserved charge is exact one. So there is no reason why
an approximately conserved charge is not generated by the unknown mechanism
at the very early universe.

We finally enumerate the remaining problems and the analyses which have not been 
carried out yet in this paper. The first problem is that we have omitted all four body
 processes such as $ll \leftrightarrow ll$ or $ll \leftrightarrow \phi \phi$, where $l$ is the
 lepton and $\phi$ is the Higgs field.
Although we expect that the rates of the  four body processes are much smaller
than those of the three body processes, the  four body processes should
be taken into account for the exactness. 
The second problem is that the analyses containing CP violation are not
 carried out.
We have dealt in the restricted parameter
 region where there are no CP violation in the lepton sector of the models in order to focus on the feature 
 that the baryon number is washed out
in the Majorana neutrino mass models. If we try to investigate the feature
more generally and more exactly, the analyses 
containing the CP violation should be carried out. 
However, even if the $SU(2)_{L}$ triplet Higgs model which has the only one 
triplet Higgs field contains CP violation in the lepton sector,
the baryon number can not be generated. 
Hence the condition to protect the primordial baryon number which
we obtain in this paper is valid in the case that the model contains the CP violation.
On the other hand, since Majorana neutrino mass matrix generally has one Kobayashi-Maskawa phase
and two Majorana phases, the analyses of the results of the neutrino oscillation 
experiments might be slightly changed.

\begin{acknowledgments}
I would like to thank C. S. Lim for helpful and fruitful discussions and for careful reading of the manuscript.  
I also thank K. Ogure for helpful discussions on this manuscript.
I am grateful to M. Kakizaki for valuable comments on this manuscript.
\end{acknowledgments}

\par

\appendix
%{\Large\bfseries Appendix}

\section{Boltzmann Equations in Sec. \ref{Washing} }
The difference between the number density of the matter and one of the anti-matter
is calculated as
\begin{eqnarray}
n_{+}-n_{-}&=&g\int \frac{d^{3}p}{(2\pi)^{3}} \biggl[ \frac{1}{e^{\frac{E-\mu}{T}}\pm 1}
-\frac{1}{e^{\frac{E+\mu}{T}}\pm 1} \biggr]  \nonumber \\
&=& \frac{1}{6} g\mu T^{2} \times
\left\{ \begin{array}{ll}
                                     1  & (\mbox{fermion}) \\
                                    2 & (\mbox{boson}). 						 
									 \end{array} \right.   \label{che}
\end{eqnarray}

We write down the  Boltzmann equation 
for the time evolution of the lepton number density $n_{l}$ as
\begin{eqnarray}
\dot{n}_{l}+3H n_{l}&=&\int d \Pi_{X} d {\Pi}_{1} d \Pi_{2} (2 \pi)^{4} \delta^{4}(p_{X}-p_{1}-p_{2})  
\nonumber \\
& \biggl[& \sum_{\alpha=e,\mu,\tau} |M_{f}(f^{\alpha \alpha})|^{2} [f(\bar{\xi}^{0})
-f(\nu_{\alpha})f(\nu_{\alpha})] +2|M_{f}(f^{e \mu})|^{2} [f(\bar{\xi}^{0})
-f(\nu_{e})f(\nu_{\mu})]  \nonumber \\
& &+2|M_{f}(f^{e \tau})|^{2} [f(\bar{\xi}^{0})
-f(\nu_{e})f(\nu_{\tau})]+2|M_{f}(f^{\mu \tau})|^{2} [f(\bar{\xi}^{0})
-f(\nu_{\mu})f(\nu_{\tau})]  \nonumber \\
\nonumber \\
&+&\sum_{\alpha=e,\mu,\tau} |M_{f}(f^{\alpha \alpha})|^{2} [f(\xi^{-})
-f(\nu_{\alpha})f(e_{\alpha L})] \nonumber  \\
& &+|M_{f}(f^{e \mu})|^{2} [f(\xi^{-})
-f(\nu_{e})f(\mu_{L})]+|M_{f}(f^{e \tau})|^{2} [f(\xi^{-})
-f(\nu_{e})f(\tau_{L})]  \nonumber \\
& &+|M_{f}(f^{\mu \tau})|^{2} [f(\xi^{-})
-f(\nu_{\mu})f(\tau_{L})]+|M_{f}(f^{e \mu})|^{2} [f(\xi^{-})
-f(e_{L})f(\nu_{\mu})]  \nonumber \\
& &+|M_{f}(f^{e \tau})|^{2} [f(\xi^{-})
-f(e_{L})f(\nu_{\tau})]+|M_{f}(f^{\mu \tau})|^{2} [f(\xi^{-})
-f(\mu_{L})f(\nu_{\tau})]  \nonumber \\
\nonumber \\
&+&\sum_{\alpha=e,\mu,\tau} |M_{f}(f^{\alpha \alpha})|^{2} [f(\xi^{--})
-f(e_{\alpha L})f(e_{\alpha L})]  \nonumber \\
& & +2|M_{f}(f^{e \mu})|^{2} [f(\xi^{--})
-f(e_{L})f(\mu_{L})]+2|M_{f}(f^{e \tau})|^{2} [f(\xi^{--})-f(e_{L})f(\tau_{L})]  \nonumber \\
& &+2|M_{f}(f^{\mu \tau})|^{2} [f(\xi^{--})
-f(\mu_{L})f(\tau_{L})]  \nonumber \\
\nonumber \\
&+&\int d \Pi_{3} \frac{\delta^{4}(p_{X}-p_{1}+p_{2}-p_{3})}{\delta^{4}(p_{X}-p_{1}+p_{2})}   \label{levo} \\
& &\sum_{\alpha=e,\mu,\tau} \frac{1}{2} |M_{s}|^{2} 
[f(\bar{u}_{L})f(\bar{u}_{L})f(\bar{d}_{L})-f(e_{\alpha L})+f(\bar{u}_{L})f(\bar{d}_{L})f(\bar{d}_{L})-f(\nu_{\alpha})
]\ \biggr].  \nonumber
\end{eqnarray}

We write down the  Boltzmann equation 
for the time evolution of the baryon number density $n_{b}$ as
\begin{eqnarray}
\dot{n}_{b}+3H n_{b}&=&\int d \Pi_{X} d {\Pi}_{1} d \Pi_{2} d \Pi_{3}
(2 \pi)^{4} \delta^{4}(p_{X}-p_{1}-p_{2}-p_{3})
\label{bevo} \\
& &\sum_{\alpha=e,\mu,\tau} \frac{1}{2} |M_{s}|^{2} 
(f(\bar{e}_{\alpha L})-f(u_{L})f(u_{L})f(d_{L})+f(\bar{\nu}_{\alpha})-f(u_{L})f(d_{L})f(d_{L})].   \nonumber  
\end{eqnarray}

We write down the  Boltzmann equation 
for the time evolution of the triplet Higgs number density $n_{\delta}$ as
\begin{eqnarray}
\dot{n}_{\delta}+3H n_{\delta}&=&  \bigcap
\nonumber \\
& \biggl[ & \sum_{\alpha=e,\mu,\tau} \frac{1}{2}|M_{f}(f^{\alpha \alpha})|^{2} [
f(\bar{\nu}_{\alpha})f(\bar{\nu}_{\alpha})-f(\xi^{0})] 
+|M_{f}(f^{e \mu})|^{2} [f(\bar{\nu}_{e})f(\bar{\nu}_{\mu}))
-f(\xi^{0})]  \nonumber \\
& &+|M_{f}(f^{e \tau})|^{2} [f(\bar{\nu}_{e})f(\bar{\nu}_{\tau})
-f(\xi^{0})]+|M_{f}(f^{\mu \tau})|^{2} [f(\bar{\nu}_{\mu})f(\bar{\nu}_{\tau})
-f(\xi^{0})]  \nonumber \\
\nonumber \\
&+&\sum_{\alpha=e,\mu,\tau}\frac{1}{2} |M_{f}(f^{\alpha \alpha})|^{2}
 [f(\bar{\nu}_{\alpha})f(\bar{e}_{\alpha L})-f(\xi^{+})] \nonumber  \\
& &+\frac{1}{2}|M_{f}(f^{e \mu})|^{2} [f(\bar{\nu}_{e})f(\bar{\mu}_{L})
-f(\xi^{+})]+\frac{1}{2}|M_{f}(f^{e \tau})|^{2} [f(\bar{\nu}_{e})f(\bar{\tau}_{L})
-f(\xi^{+})]  \nonumber \\
& &+\frac{1}{2}|M_{f}(f^{\mu \tau})|^{2} [f(\bar{\nu}_{\mu})f(\bar{\tau}_{L})
-f(\xi^{+})]+\frac{1}{2}|M_{f}(f^{e \mu})|^{2} [f(\bar{\nu}_{e})f(\bar{\mu}_{L})
-f(\xi^{+})]  \nonumber \\
& &+\frac{1}{2}|M_{f}(f^{e \tau})|^{2} [f(\bar{\nu}_{e})f(\bar{\tau}_{L})
-f(\xi^{+})]+\frac{1}{2}|M_{f}(f^{\mu \tau})|^{2} [f(\bar{\nu}_{\mu})f(\bar{\tau}_{L})
-f(\xi^{+})]  \nonumber \\
\nonumber \\
&+&\sum_{\alpha=e,\mu,\tau}\frac{1}{2} |M_{f}(f^{\alpha \alpha})|^{2} [f(\bar{e}_{\alpha L})
f(\bar{e}_{\alpha L})-f(\xi^{++})]  \nonumber \\
& &+|M_{f}(f^{e \mu})|^{2} [f(\bar{e}_{L})f(\bar{\mu}_{L})
-f(\xi^{++})]+|M_{f}(f^{e \tau})|^{2} [f(\bar{e}_{L})f(\bar{\tau}_{L})
-f(\xi^{++})]  \nonumber \\
& &+|M_{f}(f^{\mu \tau})|^{2} [f(\bar{\mu}_{L})f(\bar{\tau}_{L})
-f(\xi^{++})]  \nonumber \\
\nonumber \\
&+&\frac{1}{2}|M_{A}(A)|^{2} [f(\phi^{0})f(\phi^{0})
-f(\xi^{0})]+\frac{1}{2}|M_{A}(A)|^{2} [f(\phi^{0})f(\phi^{+})
-f(\xi^{+})]  \nonumber \\
& &+\frac{1}{2}|M_{A}(A)|^{2} [f(\phi^{+})f(\phi^{+})
-f(\xi^{++})] \ \biggr].   \label{devo}
\end{eqnarray}

We write down the  Boltzmann equation 
for the time evolution of the doublet Higgs number density $n_{\phi}$ as
\begin{eqnarray}
\dot{n}_{\phi}+3H n_{\phi}&=&  \bigcap 
\nonumber \\
& \biggl[& \sum_{i=d,s,b} |M_{d}(y_{d}^{i})|^{2}
 [f(\bar{d}_{i R})f(d_{i L})-f(\phi^{0})]  \nonumber \\
& & +\sum_{i=u,c,t} |M_{u}(y_{u}^{i})|^{2}
 [f(u_{i R})f(\bar{u}_{i L})-f(\phi^{0})]  \nonumber \\
 & & +\sum_{\alpha=e,\mu.\tau} |M_{e}(y_{e}^{\alpha})|^{2}
 [f(\bar{e}_{\alpha R})f(e_{\alpha L})-f(\phi^{0})]  \nonumber \\
 & & +\sum_{i} |M_{d}(y_{d}^{i})|^{2}
 [f(\bar{d}_{i R})f(u_{i L})-f(\phi^{+})]  \nonumber \\
 & & +\sum_{i} |M_{u}(y_{u}^{i})|^{2}
 [f(u_{i R})f(\bar{d}_{i L})-f(\phi^{+})]  \nonumber \\
& & +\sum_{\alpha=e,\mu.\tau} |M_{e}(y_{e}^{\alpha})|^{2}
 [f(\bar{e}_{\alpha R})f(\nu_{\alpha L})-f(\phi^{+})]  \nonumber \\
 \nonumber\\
 & &+|M_{A}(A)|^{2} [f(\xi^{0})-f(\phi^{0})f(\phi^{0})]  \nonumber \\
& &+|M_{A}(A)|^{2} [f(\xi^{+})-f(\phi^{0})f(\phi^{+})]  \nonumber \\
& &+|M_{A}(A)|^{2} [f(\xi^{++})-f(\phi^{+})f(\phi^{+})] \ \biggr].   \label{pevo}
\end{eqnarray}

We write down the  Boltzmann equation 
for the time evolution of the charged gauge boson number density $n_{w^{-}}$ as
\begin{eqnarray}
\dot{n}_{w^{-}}+3H n_{w^{-}}&=&  \bigcap 
\nonumber \\
& \biggl[& \sum_{i,j} |M_{g}(\frac{g}{\sqrt{2}}V_{KM}^{ij})|^{2}
 [f(\bar{u}_{i L})f(d_{j L})-f(w^{-})]  \nonumber \\
 & & +\sum_{\alpha=e,\mu.\tau} |M_{g}(\frac{g}{\sqrt{2}})|^{2}
 [f(\bar{\nu}_{\alpha})f(e_{\alpha L})-f(w^{-})]  \nonumber \\
 & & +|M_{g}(\frac{g}{\sqrt{2}}k_{\mu})|^{2}
 [f(\phi^{-})f(\phi^{0})-f(w^{-})]   \nonumber \\
 & &  +|M_{g}({g}k_{\mu})|^{2}
 [f(\xi^{-})f(\xi^{0})-f(w^{-})]  \nonumber \\
& &  +|M_{g}({g}k_{\mu})|^{2}
 [f(\xi^{--})f(\xi^{+})-f(w^{-})] \ \biggr].  \label{wevo}
\end{eqnarray}

The existences of the in-equilibrium processes make the Boltzmann equations,
 (\ref{Levo}), (\ref{Bevo}), (\ref{Devo}), (\ref{Pevo}), and (\ref{Wevo})
the simplified ones as 
\begin{eqnarray}
s\frac{dL}{dt}&=& \bigcap \cdot e^{-\frac{E}{T}} \cdot \frac{2}{T}
\nonumber \\
& \biggl[& \sum_{\alpha=e,\mu,\tau} |M_{f}(f^{\alpha \alpha})|^{2} (-3)\times
(\mu_{\xi^{0}}+2\mu_{\nu_{\alpha}})
-6|M_{f}(f^{e \mu})|^{2} (\mu_{\xi^{0}}+\mu_{\nu_{e}}+\mu_{\nu_{\mu}})  \nonumber \\
& &-6|M_{f}(f^{e \tau})|^{2} (\mu_{\xi^{0}}+\mu_{\nu_{e}}+\mu_{\nu_{\tau}})
-6|M_{f}(f^{\mu \tau})|^{2}  (\mu_{\xi^{0}}+\mu_{\nu_{\mu}}+\mu_{\nu_{\tau}})\nonumber \\
\nonumber \\
& &+\int d \Pi_{3} \frac{\delta^{4}(p_{X}-p_{1}-p_{2}-p_{3})}{\delta^{4}(p_{X}-p_{1}-p_{2})}
\nonumber \\
&& \sum_{\alpha=e,\mu,\tau}\frac{1}{2}|M_{s}|^{2} 
 (-2)\times(3\mu_{u_{L}}+\mu_{\alpha_{L}}) \ \biggr], \label{Leqb} \\
\nonumber \\
s\frac{dB}{dt}&=&\int d \Pi_{X} d {\Pi}_{1} d \Pi_{2} d \Pi_{3}
(2 \pi)^{4} \delta^{4}(p_{X}-p_{1}-p_{2}-p_{3}) \cdot e^{-\frac{E}{T}} \cdot \frac{2}{T}
\nonumber \\
& &\sum_{\alpha=e,\mu,\tau}\frac{1}{2}|M_{s}|^{2} (-2)\times
(3\mu_{u_{L}}+\mu_{\alpha_{L}}), \label{Beqb} \\
\nonumber \\
s\frac{d\Delta}{dt}&=&  \bigcap \cdot e^{-\frac{E}{T}} \cdot \frac{2}{T}
\nonumber \\
& \biggl[& \sum_{\alpha=e,\mu,\tau} |M_{f}(f^{\alpha \alpha})|^{2}(\frac{-3}{2})
(\mu_{\xi^{0}}+2\mu_{\nu_{\alpha}})-3|M_{f}(f^{e \mu})|^{2} 
(\mu_{\xi^{0}}+\mu_{\nu_{e}}+\mu_{\nu_{\mu}})  \nonumber \\
& &-3|M_{f}(f^{e \tau})|^{2} 
(\mu_{\xi^{0}}+\mu_{\nu_{e}}+\mu_{\nu_{\tau}})-3|M_{f}(f^{\mu \tau})|^{2}
(\mu_{\xi^{0}}+\mu_{\nu_{\mu}}+\mu_{\nu_{\tau}})   \nonumber \\
\nonumber \\
& &+\frac{3}{2}|M_{A}(A)|^{2} 
(2\mu_{\phi^{0}}-\mu_{\xi^{0}}) \ \biggr],  \label{Deqb}  \\
\nonumber \\
s\frac{d\Phi}{dt} &=&  \bigcap  \cdot e^{-\frac{E}{T}} \cdot \frac{2}{T}
\nonumber \\
 & &-3|M_{A}(A)|^{2} (2\mu_{\phi^{0}}-\mu_{\xi^{0}}),   \label{Peqb}  \\ 
& & \nonumber \\
W &=&  0. \label{Weqb}
\end{eqnarray}

%\begin{eqnarray}
%s\frac{dL}{dt}&=& \bigcap \cdot e^{-\frac{E}{T}} \cdot \frac{2}{T}
%\nonumber \\
%& & (-27) \times |M_{f}(f)|^{2} (\mu_{\xi^{0}}+\frac{2}{3}\mu_{\nu})  \nonumber \\
%\nonumber \\
%& &+\int d \Pi_{3} \frac{\delta^{4}(p_{X}-p_{1}-p_{2}-p_{3})}{\delta^{4}(p_{X}-p_{1}-p_{2})} 
%\frac{1}{2}|M_{s}|^{2} (-2)\times(9\mu_{u_{L}}+\mu_{L}) \label{Lfb}  \\
%\nonumber \\
%s\frac{dB}{dt}&=& \bigcap \cdot e^{-\frac{E}{T}} \cdot \frac{2}{T}   \nonumber \\
%& &+\int d \Pi_{3} \frac{\delta^{4}(p_{X}-p_{1}-p_{2}-p_{3})}{\delta^{4}(p_{X}-p_{1}-p_{2})} 
%\frac{1}{2}|M_{s}|^{2} (-2)\times(9\mu_{u_{L}}+\mu_{L})  \label{Bfb} \\
%\nonumber \\
%s\frac{d\Delta}{dt}&=&  \bigcap \cdot e^{-\frac{E}{T}} \cdot \frac{2}{T}
%\nonumber \\
%& & (-\frac{27}{2}) |M_{f}(f)|^{2}
%(\mu_{\xi^{0}}+\frac{2}{3}\mu_{\nu})  \nonumber \\
%\nonumber \\
%& &+\frac{3}{2}|M_{A}(A)|^{2} 
%(2\mu_{\phi^{0}}-\mu_{\xi^{0}})   \label{Dfb} \\
%\nonumber \\
%s\frac{d\Phi}{dt} &=&  \bigcap  \cdot e^{-\frac{E}{T}} \cdot \frac{2}{T}
%\nonumber \\
% & &-3|M_{A}(A)|^{2} (2\mu_{\phi^{0}}-\mu_{\xi^{0}})  \label{Pfb}
%\end{eqnarray}

\end{document}